\documentclass[preprint]{aastex62}
\usepackage{amsmath}
\submitjournal{ApJ}

\shorttitle{Obliquity Evolution of Circumstellar Planets in Sun-like Stellar Binaries}
\shortauthors{Quarles et al.}

\newcommand{\degree}{$^\circ$}
\begin{document}

\title{Obliquity Evolution of Circumstellar Planets in Sun-like Stellar Binaries}

\correspondingauthor{Billy Quarles}
\email{billylquarles@gmail.com}

\author[0000-0002-9644-8330]{Billy Quarles}
\affil{Center for Relativistic Astrophysics, School of Physics, 
Georgia Institute of Technology, 
Atlanta, GA 30332, USA}

\author{Gongjie Li}
\affil{Center for Relativistic Astrophysics, School of Physics, 
Georgia Institute of Technology, 
Atlanta, GA 30332, USA}

\author{Jack J. Lissauer}
\affil{NASA Ames Research Center, 
Space Science and Astrobiology Division, MS 245-3, 
Moffett Field, CA 94035, USA}



\begin{abstract}
Changes in planetary obliquity, or axial tilt, influence the climates on Earth-like planets.  In the solar system, the Earth's obliquity is stabilized due to interactions with our moon and the resulting {small amplitude variations ($\sim$2.4\degree)} are beneficial for advanced life.  Most Sun-like stars have at least one stellar companion and the habitability of circumstellar exoplanets is shaped by their stellar companion.  We show that a stellar companion can dramatically change whether {Earth-like obliquity stability is} possible through planetary orbital precession relative to the binary orbit or resonant pumping of the obliquity through spin-orbit interactions.  We present a new formalism for the planetary spin precession that accounts for orbital misalignments between the planet and binary.  Using numerical modeling in $\alpha$ Centauri AB we show: a stark contrast between the planetary obliquity variations depending on the host star, planetary neighbors limit the possible spin states for {Earth-like obliquity stability}, and the presence of a moon can destabilize the obliquity, defying our Earth-based expectations.  An Earth-like rotator orbiting the primary star would experience {small} obliquity variations for 87\%, 74\%, or 54\% of Solar type binaries, depending on the mass of the primary (0.8, 1.0, or 1.2 M$_\odot$, respectively).  Thus, Earth-like planets likely experience much larger obliquity variations, with more extreme climates, unless they are in specific states, such as orbiting nearly planar with the binary and rotating retrograde (backwards) like Venus.

\end{abstract}



\section{Introduction} \label{sec:intro}
The Earth's obliquity, or axial tilt $\epsilon$, changes slowly with time, with an oscillation cycle from 22.1\degree -- 24.5\degree{} completing in $\sim$41,000 years.  Additionally, the other solar system planets induce changes in Earth's orbit through gravitational interactions over $\sim$76,000 years.  The value of Earth's obliquity impacts the seasonal cycles and the long-term variation of obliquity affects the planetary climate \citep{Milankovictch1969}, which has been deduced from the geologic record \citep{Kerr1987,Mitrovica1995,Pais1999}.  This benign obliquity variation ($\Delta\epsilon \leq 2.4$\degree) for the Earth is modulated by interactions with the Moon, or Luna, \citep{Laskar1993b}, where the variation would have been larger for a Moonless Earth \citep{Laskar1993b,Lissauer2012,Li2014}.  Moreover, the obliquity of the  solar system's other terrestrial planets can be chaotic under the right conditions \citep{Laskar1993c} leading to dramatically different climates over their history.

Mars does not experience a strong stabilization from its moons, both of which are very small.  Mars's obliquity variation has reached $\sim$60\degree{} \citep{Ward1973,Touma1993}, which contributed to its atmospheric collapse \citep{Head2003,Head2005,Forget2013} in addition to processes that alter the atmospheric pressure (i.e., atmospheric erosion due to stellar winds \cite{Mansfield2018,Kite2019}).  If Venus had a thin atmosphere and rotated rapidly (with a period comparable to those of Earth and Mars at present), its obliquity variations due to perturbations from Jupiter would have been large \citep{Barnes2016}, however a combination of effects (atmospheric tides \& core-mantle friction) have likely slowed its spin \citep{Correia2003}.  Even the spin-orbit interactions between the outer giant planets have affected their obliquities \citep{Ward2004}.  Recently, detailed numerical studies were performed for compact exoplanetary systems (Kepler-62 \& Kepler-186) showing that {Earth-like (i.e., small)} obliquity variations are possible for terrestrial planets within their respective habitable zones \citep{Deitrick2018,Shan2018}, where larger variations could be induced for Kepler-62f through spin-orbit interactions if long period ($>1000$ days) giant planets exist within the system \citep{Quarles2017}.  Investigations of obliquity variation for nearly coplanar terrestrial planets in binary systems have uncovered that orbital precession induced from the stellar companion can leave an imprint on planetary climates akin to Milankovitch cycles \citep{Forgan2012,Forgan2016}.

Stellar surveys that focus on multiplicity showed that most Sun-like stars have a stellar companion with a modest eccentricity ($e\lesssim0.4$) and a twin rate of $\sim$10\% \citep{Raghavan2010,Moe2017}.  Moreover, polarimetric observations of disks in binaries show a $\sim$10\degree{} misalignment of the disk with the binary plane \citep{Monin2006}, which suggests that the planets that form may be inclined.  Current observations by the Transiting Exoplanet Survey Satellite (TESS) are expected to uncover $\sim$500,000 eclipsing binary systems \citep{Sullivan2015}, where a subset of those are expected to host planets.  Indeed, circumstellar planets in binaries were identified within the K2 data (K2-136 \cite{Ciardi2018}; K2-288 \cite{Feinstein2019}), which observed a much smaller number of stars.  The two largest stars in the nearest stellar system to us, $\alpha$ Centauri AB ($\alpha$ Cen AB), are Sun-like in terms of their mass and radius \citep{Pourbaix2016}.  Detection of confirmed planets in this binary system have been elusive due to false positives produced in analyzing large archives of observations \citep{Hatzes2013,Rajpaul2016}, but efforts are underway to observe any potentially habitable planets in residence through direct imaging \citep{Bendek2015}.

We seek to better understand the general outcomes of obliquity evolution for exoplanets orbiting either star in a Sun-like binary system and the possible implications for planet habitability through radiative flux variations.  To identify the possibilities, we investigate the dynamical evolution of planets within the habitable zone of either star in $\alpha$ Cen AB and generalize our results to the larger binary star population.  Our methods, the initial conditions for our n-body simulations, the numerical setup for evaluating the spin evolution, and the background necessary for calculating the radiative flux for a planet are summarized in Section \ref{sec:methods}. In Section \ref{sec:results}, we present the results of our numerical simulations considering a single planet orbiting $\alpha$ Cen B, a discussion of the effects due to additional bodies (neighboring planets and a moon) accompanying the Earth-like world around either star in $\alpha$ Cen AB, and a generalization to a wide range of binary orbital architectures that may be observed with TESS.  We provide the summary and conclusions of our results with comparisons to previous studies in Section \ref{sec:conclusions}.    

\section{Methods} \label{sec:methods}
\subsection{n-body Simulations} \label{sec:nbody}
In order to identify the orbital variations over long timescales, we use the \texttt{mercury6} integrator that has been modified to efficiently evaluate planetary orbits within binary star systems \citep{Chambers2002}.  When evaluating planetary orbits using the $\alpha$ Cen binary, {the binary starts} with the latest stellar parameters (e.g., stellar masses, semimajor axis, eccentricity, etc.) determined through observations \citep{Pourbaix2016}, where these values are listed in Table 1 of \cite{Quarles2018a}.  The simulations are performed relative to the plane of the binary, and thus we set the binary inclination to 0\degree{} and equate the binary argument of periastron to the observed longitude of periastron ($\omega_{bin} = \varpi_{obs}$).  For our more general simulations of binary systems, we vary the mass ratio $\mu$ (=M$_B$/(M$_A$+M$_B$)) of the binary, the semimajor axis $a_{bin}$, and the eccentricity $e_{bin}$.  In these runs, the argument of periastron for the binary begins equal to 0\degree{} and the mean anomaly starts at apastron ($MA_{bin}=180$\degree).

The binary system along with the planet(s) are evolved for 50 Myr, which is sufficient for slow changes to the planetary orbit to develop (e.g., Kozai-Lidov (KL) oscillations \citep{Kozai1962,Lidov1962}).  A simulation is terminated if the radial distance from the binary center of mass exceeds 500 AU {(twice the largest binary semimajor axis considered)} or a collision occurs.  These conditions are never met for our nearly planar simulations using $\alpha$ Cen AB, but do occur for some initial conditions when the planetary inclination or binary eccentricity are sufficiently large.  For each simulation, we record the system state every 100 years and use this output in another routine that calculates the obliquity evolution (see Section \ref{sec:spin}).

\subsection{Spin Evolution of a Planet} \label{sec:spin}
The obliquity $\epsilon$ of a planet is defined by the angle between its spin momentum vector $S_p$ and orbital angular momentum vector $L_p$ \citep[][see their Figure 1]{Shan2018}.  The vector $S_p$ can also rotate about $L_p$, where the rotation angle defines the planetary spin longitude $\psi$. The spin precession $\dot{\psi}$ (i.e., rotation rate of the spin vector) for an Earth-like planet is due to torques exerted on the equatorial bulge from the host star.  The spin precession constant $\alpha$ (in $\arcsec$/yr) is a parameter that depends on the planet's rotation (flattening and rotation period;  \cite{Laskar1993b,Li2014,Quarles2017}).  The magnitude of the spin precession varies with the cosine of the obliquity ($\dot{\psi} = \alpha \cos \epsilon$).  Most of our simulations evaluate a grid of initial values, unless otherwise noted, for the spin precession constant ($\alpha \leq 100$ \arcsec/yr) and obliquity ($0^\circ \leq \epsilon_o < 90^\circ$) using 1 \arcsec/yr and 1\degree{} increments, respectively.

Our model uses a constant spin period because the timescale for significant increases in spin period due to tidal interactions with the host star is longer than our integration timescale.  As a result, we make use of the secular time-dependant Hamiltonian that includes the canonical variable $\chi$ ($=\cos \epsilon$) and spin longitude $\psi$ in the following equations of motion \citep{deSurgy1997,Saillenfest2019}:

\begin{align} \label{eqn:eom}
    \frac{\delta \psi}{\delta t} &= \frac{\alpha \chi}{\left(1-e^2\right)^{3/2}} - \frac{\chi}{\sqrt{1-\chi^2}}\left[ \mathcal{A}(t)\sin \psi + \mathcal{B}(t) \cos \psi \right] - 2\mathcal{C}(t)  \\ 
    \frac{\delta \chi}{\delta t} &= \sqrt{1-\chi^2}\left[\mathcal{B}(t)\sin\psi - \mathcal{A}(t)\cos \psi \right], 
\end{align}
\noindent where the functions $\mathcal{A}(t)$, $\mathcal{B}(t)$, and $\mathcal{C}(t)$ depend on the orbital evolution of the planet through $p = \sin(i/2)\sin \Omega$ and $q = \sin(i/2)\cos \Omega$ in the following relations:

\begin{align}
    \mathcal{A}(t) &= 2\left(\dot{q}+p\left(q\dot{p} - p\dot{q} \right) \right)/\sqrt{1-p^2-q^2},  \\
    \mathcal{B}(t) &= 2\left(\dot{p}-q\left(q\dot{p} - p\dot{q} \right) \right)/\sqrt{1-p^2-q^2}, \\
    \mathcal{C}(t) &= \left(q\dot{p} - p\dot{q} \right).
\end{align}

The obliquity was evolved using the numerical integration routines from the \texttt{scipy} library \citep{Jones2001} within \texttt{python} in decade steps between each state that is recorded from a given n-body simulation (see Section \ref{sec:nbody}).  The major effects on the spin dynamics of a planet in a binary system arise from the n-body perturbations on the planet's orbit from the companion star on a secular timescale, while the host star induces a precession on the planet’s spin.  The secondary star has a negligible effect on changing the planetary spin precession due to its distance and the relatively short interaction time during close approaches.  We also compared this approach to a more computationally expensive method using \texttt{smercury}, which is a different modified version of \texttt{mercury6} that implements the rigid body equations of motion coupled with the n-body evolution \citep{Touma1994,Lissauer2012} and found that the range of obliquity variation $\Delta \epsilon$ ($=\epsilon_{max} - \epsilon_{min}$) to be consistent between the two methods.

\begin{figure}
    \centering
    \includegraphics[width=\linewidth]{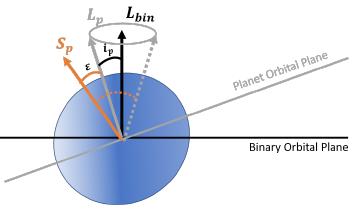}
    \caption{Schematic defining the planetary obliquity $\epsilon$ and inclination $i_p$ relative to the orbital planes of the planet ($L_p$) and binary ($L_{bin}$).  The obliquity is defined by the angle $\epsilon$ between $L_p$ and the spin angular momentum vector $S_p$.  The vector $L_p$ can precess about $L_{bin}$ at a different rate than $S_p$ does about $L_p$ resulting in a maximum obliquity, $\epsilon_{max} \approx \epsilon + 2i_p$.  When the precession rates of $S_p$ and $L_p$ are similar, spin-orbit resonances force the maximum obliquity to higher values.}
    \label{fig:spin_vector}
\end{figure}

The planetary orbital plane can be significantly inclined relative to the binary plane due to a misaligned protoplanetary disk \citep{Monin2006} or by planet-planet scattering after the disk dissipates \citep{Quintana2002}.  The system of vectors ($S_p$ \& $L_p$) are now shifted by the planetary inclination $i_p$, which is the angle between the orbital angular momentum of the planet and the binary (see Figure \ref{fig:spin_vector}).  Additionally, the planetary orbital angular momentum $L_p$ will precess over time with a characteristic frequency (i.e., nodal precession).  This orbital precession frequency, $g_s$ in {units of} radians/yr, is calculated via the Laplace-Lagrange equations \citep{Brouwer1961,Kaula1962,Heppenheimer1978,Andrade2016} up to $O(e_p^2)$:  

\begin{equation} \label{eqn:sec_freq}
    g_s = \frac{3}{4}\frac{M_{B/A}}{M_{A/B}} \left(\frac{a_p}{a_{bin}}\right)^3 \sqrt{\frac{G(M_{A/B}+M_p)}{a_p^3(1-e_{bin}^2)^3}},
\end{equation}

\noindent where $g_s$ depends on the mass of the binary stars ($M_A$ \& $M_B$ in {units of} M$_\odot$), the binary semimajor axis ($a_{bin}$ in {units of} AU), the binary eccentricity ($e_{bin}$), the planetary semimajor axis ($a_p$ in AU), and to a much lesser degree the planetary mass ($M_p$ in {units of} M$_\odot$).  In order to obtain a frequency in radians/yr the gravitational constant $G$ is $4\pi^2$, and then $g_s$ can be converted to $\arcsec$/yr when necessary.  As the planetary ascending node ($\Omega_p$) precesses due to the stellar companion, the obliquity evolution is affected through the spin longitude $\psi$, which allows the obliquity to increase by twice the planetary inclination $i_p$.  However, some initial conditions allow for the spin longitude and ascending node of a planet to precess at the same rate (i.e., $g_s = \alpha \cos \epsilon$) in a spin-orbit resonance, where this can be derived using a simplified Hamiltonian approach for small planetary inclination \citep{Li2014}.  In the case of circumstellar planets in binary systems, the planetary inclination $i_p$ is likely to be non-negligible and thus we determine, using the condition $\psi = \Omega_p$, a modified relation for the spin precession constant $\alpha$ as:

\begin{equation} \label{eqn:alpha_prec}
    \alpha_{\psi=\Omega_p} = \frac{g_s}{\cos \epsilon} - \frac{g_s \sin i_p}{\sin \epsilon},
\end{equation}

\noindent defining a fixed point in the phase space of variables $\epsilon \cos \Phi$ and $\epsilon \sin \Phi$, where $\Phi = \Omega_p - \psi$.  At the fixed point, the obliquity variation is minimized because the $S_p$ and $L_p$ vectors are more tightly coupled as they precess together and maintain a nearly constant angle between them.  The spin precession can still drive obliquity variations through the planetary eccentricity variation (see Equation \ref{eqn:eom}), but the eccentricity variation can be minimized by choosing initial conditions near the forced eccentricity $e_F$ \citep{Quarles2018a}.

\subsection{Radiative Flux on a Planet} \label{sec:flux_calc}
The radiative flux that a planet receives at the top of the atmosphere has been used as a proxy for potential climatic conditions in works for single stars \citep{Armstrong2014,Kane2017,Deitrick2018,Quarles2017}.  We employ a calculation of surface flux for a planet orbiting $\alpha$ Cen B using a similar approach.  The flux at a given orbital distance decreases rapidly with distance ($S \propto L_\star/r^2$), and the radiative contribution of $\alpha$ Cen A at its pericenter is only at most a few percent.  Moreover, the binary  spends most of its orbit far from this special location \citep{Quarles2018b}.  In addition, the planet’s position is quickly changing relative to $\alpha$ Cen A, so  the radiative contribution of $\alpha$ Cen A is small.

Following previous works \citep[i.e.,][]{Armstrong2014,Kane2017}, our calculation of the surface flux depends on the half-angle of daylight $\eta$, the latitude $\delta$, the obliquity $\epsilon$, the stellar longitude $L_s$,  and the substellar latitude $\delta_\star (= \epsilon \sin L_s)$.  The stellar longitude relates to the relative orbital phase when the northern hemisphere vernal equinox occurs.  Our analysis involves the average flux as a function of latitude over a complete orbit and we choose 90\degree{} for simplicity so that $L_s=90^\circ+f$, where $f$ denotes the planetary true anomaly.  Combining these parameters, we calculate the latitudinal flux $I_d$, or insolation, through the following:

\begin{equation}
    I_d = \frac{L_\star}{4\pi^2 r^2}\left[ \eta \sin \delta \sin \delta_\star + \sin \eta \cos \delta \cos \delta_\star \right],
\end{equation}

\noindent where the radial distance from the host star, $r$, is determined through the two-body solution of the Kepler problem at a given epoch and $L_\star$ represents the stellar luminosity of the host star \citep{Armstrong2014,Forgan2012,Forgan2016,Quarles2017,Kane2017}.  The luminosity of $\alpha$ Cen A and $\alpha$ Cen B is 1.519 L$_\odot$ and 0.5 L$_\odot$, respectively.  The average flux the planet receives per orbit, $F_{avg}$, is calculated by summing the latitudinal flux $I_d$ over the true anomaly $f$ along with a weight ($dM/df$) that accounts for the relative time spent at each phase throughout the orbit.  We also determine the change in flux $\Delta F$, or flux variation, for a given latitude over an orbit through the difference between the highest and lowest attained flux values.  Apart from flux variations due to obliquity, there can also be differences in the flux variation over time due to changes in planetary eccentricity.  In order to mitigate this effect for a \emph{single} planet orbiting $\alpha$ Cen B with $a_p = \sqrt{0.5}$ AU, we begin the Earth-like planet near the forced eccentricity $e_F \approx 0.0257$ \citep{Quarles2018a}.  Our other simulations that include additional bodies, the Earth-like planet begins on a nearly circular, coplanar orbit.

\subsection{Single Planets within Stellar Binaries} \label{sec:meth_binprec}

In order to quantify the effect on planetary obliquity due to a stellar companion, we numerically evolve the obliquity of a \emph{single} Earth-mass planet orbiting either stellar host in a binary system at the inner edge of the conservative habitable zone where the planet receives an Earth-like equivalent flux ($a_p = \sqrt{L_\star/S_\oplus}$).  The initial planetary orbit is prescribed relative to the binary orbital plane ($i_{bin}=\Omega_{bin}=0^\circ$), and the mutual inclination of the planetary orbit $i_p$ is defined by the misalignment from the binary.  The initial planetary ascending node $\Omega_p$ for these simulations is aligned with the binary orbital plane (i.e., $\Omega_p = \Omega_{bin} = 0^\circ$).  Using this setup, we perform two sets of simulations surveying more general binaries with single planets that begin on nearly circular orbits ($e_p = 10^{-6}$).  

\subsubsection{Effects From Varying Planetary Inclination and Binary Eccentricity} \label{sec:bin_set1}
 In \emph{Set 1}, we explore variations in the mass ratio $\mu$ (0.3 and 0.5), the binary semimajor axis $a_{bin}$ (20 AU and 100 AU), the binary eccentricity $e_{bin}$ (0.1, 0.3, 0.5, 0.7, and 0.9), and the planetary inclination $i_p$ ($0^\circ-180^\circ$) {as summarized in Table \ref{tab:Set1}}.  In order to minimize the changes in planetary semimajor axis due to a changing habitable zone when $\mu=0.3$, the mass of Star A is Sun-like (1 M$_\odot$) when it is the host star and the mass of Star B is slightly smaller (0.8 M$_\odot$) when it hosts the planet.  The dynamical mass ratio ($\mu = M_B/(M_A + M_B)$) determines the mass of the other star and the luminosity of Star B is determined through the mass-luminosity relation for Sun-like stars ($L\propto M^4$).  We sample the planetary inclination differently between two regimes due to the large changes in the planetary inclination through the KL mechanism \citep{Kozai1962,Lidov1962,Naoz16}.  For $i_p\leq 30^\circ$ or $i_p\geq 150^\circ$, we take 2\degree{} steps in the planetary inclination and in the intermediate regime ($40^\circ \geq i_p \geq 140^\circ$), 10\degree{} steps are taken in the planetary inclination because of the large fluctuations of the angular momentum vector $L_p$.  Moreover, we use the Bulirsch-Stoer integration scheme to ensure the accuracy of our n-body integrations when the planetary eccentricity oscillates to high values ($e_p \gtrsim 0.8$).
 
 \begin{deluxetable}{lc}
\tablecaption{Parameters Explored in \emph{Set 1} \label{tab:Set1}}
\tablehead{ \colhead{Parameter} & \colhead{Value}}
\startdata
$\mu$ & 0.3, 0.5 \\
$e_{bin}$ & 0.1, 0.3, $\ldots$, 0.9 \\
$a_{bin}$ & 20, 100 AU \\
$i_p^{pro}$ & $0^\circ$, $2^\circ$, $\ldots$, $30^\circ$ \\
$i_p^{KL}$ & $30^\circ$, $40^\circ$, $\ldots$, $140^\circ$ \\
$i_p^{ret}$ & $140^\circ$, $142^\circ$, $\ldots$, $180^\circ$
\enddata
\tablecomments{{The stepsize in the planetary inclination $i_p$ is $2^\circ$ within regimes where the Kozai-Lidov (KL) mechanism is negligible (e.g., $i_p^{pro}$ and $i_p^{ret}$), but increases to $10^\circ$ when the KL mechanism is strong ($i_p^{KL}$).}}
\end{deluxetable}

For each choice of these stellar and planetary parameters, we evaluate a coarse grid in terms of the spin precession $\alpha$ from $0-100 \arcsec$/yr (with 5 $\arcsec$/yr steps) and the initial obliquity $\epsilon_o$ from $0^\circ-180^\circ$ (with 5\degree{} steps).  From each grid, we determine the median obliquity $\left<\Delta \epsilon \right>$ attained because previous investigations of the solar system and exoplanets orbiting single stars \citep{Laskar1993b,Li2014,Quarles2017,Shan2018} have shown that regions of the $\alpha$ vs. $\epsilon_o$ phase space can have significant obliquity variation near spin-orbit resonance or resonance overlap, while the remainder (sometimes majority) of the phase space typically undergoes obliquity variation due to the precession of the orbit.  When the planetary obliquity varies primarily due to orbital precession, we expect the obliquity variation to be nearly equal to twice the initial planetary inclination ($\Delta \epsilon \approx 2i_p$).

\subsubsection{Effects From Varying Binary Semimajor Axis and Eccentricity} \label{sec:bin_set2}
In \emph{Set 2}, we explored variations in the shape of the binary orbit and kept the mass ratio similar to the $\alpha$ Cen AB system ($\mu \approx 0.46$) due to the higher abundance of stellar twins from observational studies of binary stars \citep{Raghavan2010,Moe2017}.  The Earth-like rotator ($P_{rot}=23.934$ hr \& $\epsilon_o=23.4^\circ$) initially orbits the more massive primary (Star A) at the inner edge of the conservative habitable zone and is inclined ($i_p = 5^\circ$) relative to the binary orbital plane.  The binary orbit semimajor axis $a_{bin}$ ($4-200$ AU in 1 AU steps) and eccentricity $e_{bin}$ ($0-0.9$ in 0.01 steps) are varied, where we investigate the possible binary orbits that maximize the obliquity variations.  These simulations were limited to 20 million years, which corresponds to $\sim$15 secular cycles for binaries with $a_{bin}=200$ AU.  

\subsection{Multiple Planets and a Moon in $\alpha$ Cen AB} \label{sec:meth_mult}
In the solar system, the Earth's orbit is affected by perturbations from Jupiter, the other terrestrial planets, and our moon, Luna.  These perturbations combine to increase Earth's spin precession frequency and in case of our moon, stabilize the Earth's obliquity \citep{Laskar1993a,Laskar1993b}.  We explore whether a similar outcome would be possible for an Earth-like planet orbiting either star in $\alpha$ Cen AB.  For these simulations, terrestrial planet analogs (Mercury, Venus, Earth, \& Mars) are added around either star, where the Earth-like planet's initial semimajor axis lies at the inner edge of the conservative habitable zone.  The initial semimajor axes of the remaining planets were scaled so that the dynamical spacing (measured in units of mutual Hill spheres \cite{Chambers1996,Quarles2018b}) between the terrestrial planets is preserved (see Table \ref{tab:SS_analogs}).  

\begin{deluxetable}{lcccccccc}
\tablecaption{Table of Initial Conditions for the Solar System and Analogs in $\alpha$ Centauri AB \label{tab:SS_analogs}}
\tablehead{ \colhead{} & \colhead{Sun}& \colhead{$\alpha$ Cen A} & \colhead{$\alpha$ Cen B} &&&&&\\ \colhead{Planet} & \colhead{$a$ (AU)} & \colhead{$a$ (AU)} & \colhead{$a$ (AU)} & \colhead{$e$}  & \colhead{$i$ (deg.)}  & \colhead{$\omega$ (deg.)}  & \colhead{$\Omega$ (deg.)}  & \colhead{$MA$ (deg.)} }
\startdata
Mercury & 0.38709821 & 0.49680871 & 0.27113825 & 0.20563029 & 7.0050141 & 29.124283 & 48.330537 & 174.79588 \\
Venus & 0.72332667 & 0.90353638 & 0.50987156 & 0.00675564 & 3.3945898 & 55.183460 & 76.678386 & 50.117254 \\
Earth & 0.99999293 & 1.2324772 & 0.70710678 & 0.01669905 & 0.000110 & 322.30828 & 141.22276 & 358.45157\\
Mars & 1.5236892 & 1.8450718 & 1.0818742 & 0.09331972 & 1.8498785 & 287.46216 & 49.561895 & 19.355916\\
\enddata
\tablecomments{Initial orbital elements used in our simulations of multiple planet systems orbiting each star in $\alpha$ Centauri AB. The semimajor axis values are scaled so that the third planet receives an Earth equivalent of radiative flux and the other elements are taken from \cite{Murray1999}.  Additionally the orbital elements for our outer gas giants can be found in \cite{Murray1999}.  We use initial conditions for the binary orbit from \cite{Pourbaix2016}, where detailed values can be found in \cite{Quarles2018a}.}
\end{deluxetable}

 A second set of simulations explore the added spin precession on an Earth-like planet due to a moon.  In these simulations, we do not include the possible tidal interactions.  We did perform a separate set of simulations using a rigid body integrator, \texttt{smercury} \citep{Lissauer2012}, and found consistent results with our secular approach to the obliquity evolution.  For consistency, we use the secular method throughout the remaining simulations.  Through our secular method, the added precession due to a moon modifies the spin precession constant $\alpha$ (radians/yr) to include the mean motion of the satellite through the following:
 \begin{equation} \label{eqn:moon_prec}
 \alpha = \frac{3GJ_2}{2\nu \bar{C}} \left( \frac{M_{A/B}}{a^3} + \frac{m_{moon}}{a_{moon}^3} \right ),
 \end{equation}
\noindent where the parameters are defined as the rotation rate $\nu$, the Gravitational Constant $G$, the dynamical oblateness $J_2$, moment of inertia coefficient $\bar{C}$, mass of the host star M$_{A/B}$, semimajor axis of the planet $a$, mass of the moon $m_{moon}$, and semimajor axis of the moon $a_{moon}$ relative to the host planet.  Equation \ref{eqn:moon_prec} could be more complete if we included the eccentricity of the planet and moon as well as the orbital inclination of the moon relative to the planetary equatorial plane \citep{Li2014}, but the changes in the constant $\alpha$ would be small.  To connect the changes in spin precession to physical parameters of moons, we perform additional numerical simulations using the terrestrial analogs that orbit $\alpha$ Centauri B, where the mass in M$_{luna}$ (= 1/81 M$_\oplus$) and semimajor axis  $a_{luna}$ (=0.00257 AU) of the moon are varied from $0.0001-2$  M$_{luna}$ using 1000 evenly spaced samples on a base-10 logarithmic scale and 1000 evenly spaced samples from $0.1-1.3$ $a_{luna}$ on a linear scale.  The Earth-mass planet begins with an Earth-like obliquity ($\epsilon_o=23.4^\circ$) and spin period ($P_{rot}=23.934$ hr).

\subsection{Binary Star Populations} \label{sec:meth_binpop}
Observational studies of binary stars \citep{Raghavan2010,Moe2017} have produced statistical distributions for the occurrence of binaries that depend on the mass of the primary M$_A$, the orbital period (or semimajor axis), the mass quotient $q$ ($=M_B/M_A$), and the eccentricity $e_{bin}$.  We use these results to build a probability distribution (PDF) for the stellar mass quotient $q$ ranging from $0.1-1$, where we adjusted the PDF for the range from $0.95-1$ to account for the occurrence of stellar twins.  We use the results from \cite{Moe2017} that provide a joint PDF that accounts for the occurrence for the orbital period with a log{-}normal distribution and power law distributions for the binary mass quotient {($p_q \propto q^\gamma$)} and eccentricity {($p_e \propto e^\eta$)}.

\begin{figure}
    \centering
    \includegraphics[width=\linewidth]{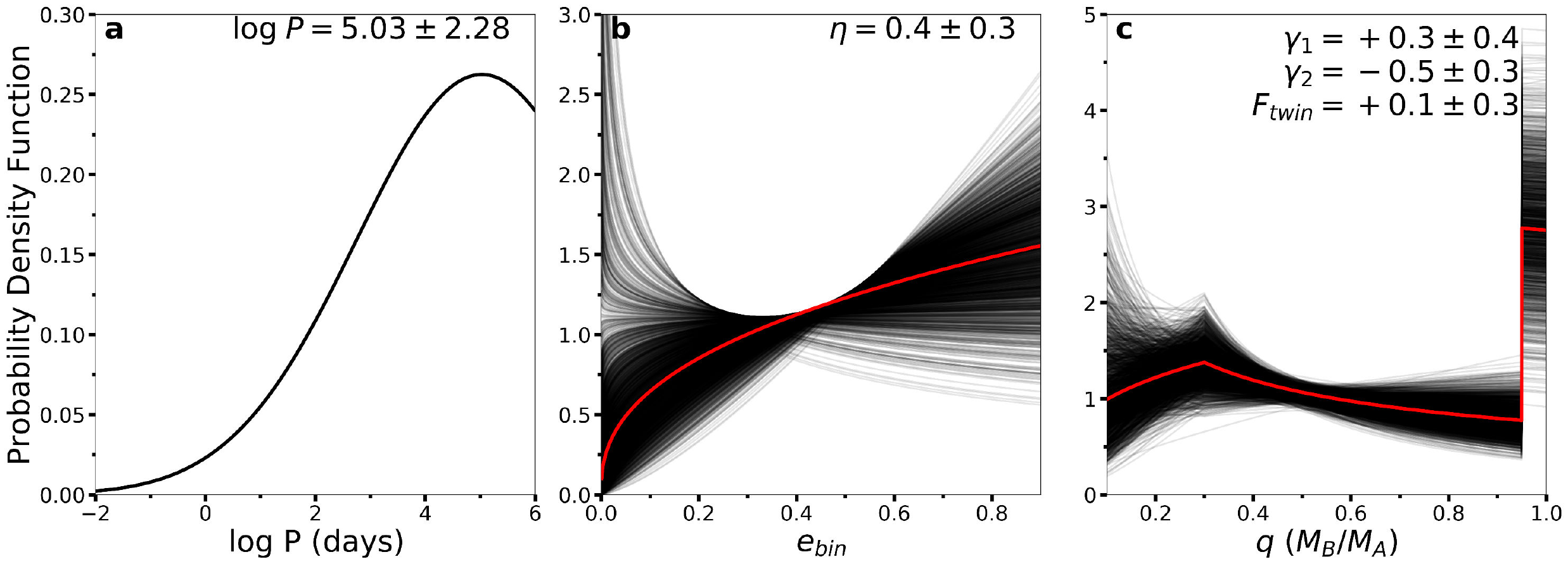}
    \caption{{Probability distribution functions (PDFs) derived from binary star population studies \citep{Raghavan2010,Moe2017} for the binary (a) period, (b) eccentricity, and (c) mass quotient used in \emph{Set 2}.  The PDF for the binary period is a log-normal distribution, where the eccentricity and mass quotient follow power laws (e.g., $e^\eta$ and $q^\gamma$).  The mass quotient is further qualified by a broken power law for two domains: $\gamma_1$ for $0.1\leq q \leq 0.3$ and $\gamma_2$ for $0.3 < q \leq 1$.  There is also an excess fraction of nearly equal mass binaries ($0.95\leq q \leq 1$) represented by $F_{twin}$ (see Fig. 2 in \cite{Moe2017}).  The red curves in (b) and (c) represent an example distribution using the median value, where the black curves are samples assuming Gaussian errors in the respective parameters.} }
    \label{fig:bin_pdf}
\end{figure}

Our general simulations for \emph{Set 2} focus on binary systems ($e_{bin}\leq0.9$; $a_{bin}\leq250$ AU) where a planet could stably orbit around a Solar-Type primary and the perturbations from the secondary could potentially influence the planet's obliquity.  As a result, we sample binary orbital periods ($\log\;P\leq 6$) using a mean value of 5.03, a standard deviation of 2.28, and a normalization of 1.5 \citep{Raghavan2010}.  When applying the PDF for the binary mass quotient, we use power law distributions and considered 2 possible domains: $0.1\leq q\leq1$ and $0.3\leq q\leq1$.  The occurrence rates are not as well known for $0.1\leq q \leq 0.3$, so we use a broken power law and enforce continuity across the boundary at q=0.3 \citep{Moe2017}.  For very low mass ratios $0.1\leq q \leq 0.3$, the power law slope {$\gamma_1 = 0.3\pm0.4$}.  For the other range $0.3\leq q \leq 1$, the power law slope {$\gamma_2 = -0.5\pm0.3$}.  Lastly, we use a single power law to estimate the PDF for the binary eccentricity, where the slope of the power law  {$\eta = 0.4\pm0.3$}.  Each of the power laws were normalized (within the respective domain) so that the sum of the probability for the area within the entire domain is equal to unity.  {Figure \ref{fig:bin_pdf} illustrates the shape of our PDFs using the above prescribed values, as well as how the PDF varies with respect to the uncertainty in the parameters for the power laws.}  The standard method of Monte Carlo integration is performed using these observationally derived PDFs, where we use 250 million samples to produce our final estimates.

\section{Results and Discussion} \label{sec:results}

\subsection{A \emph{Single} Inclined Earth-mass Planet Orbiting $\alpha$ Centauri B} \label{sec:single_aCenB}
In $\alpha$ Cen AB, changes in the planetary orbit can arise from stellar perturbations \citep{Quarles2016,Quarles2018a}, and these changes can alter the obliquity of the planet on astrobiological timescales.  For an Earth-mass planet orbiting either star $\alpha$ Cen A or B at the inner edge of the conservative habitable zone, the obliquity variation $\Delta \epsilon$ ($=\epsilon_{max} - \epsilon_{min}$) depends on the assumed spin precession constant $\alpha$ induced by the host star, initial obliquity $\epsilon_o$, and initial spin longitude $\psi_o$ (see Section \ref{sec:spin}).  We investigate first the influence of these parameters ($\alpha$, $\epsilon_o$, \& $\psi_o$) on a \emph{prograde-spinning} Earth-mass planet orbiting $\alpha$ Cen B.  In Section \ref{sec:gen_bin}, we will discuss results considering a similar planet orbiting $\alpha$ Cen A and a wide range of similar binaries.

Before starting our numerical simulations, we identify the secular frequency ($g_s \approx 76.63$ \arcsec/yr; see Equation \ref{eqn:sec_freq}) for which we expect the orbit of an Earth-mass planet to oscillate using the orbital elements of the binary and our starting planetary semimajor axis ($a_\oplus = \sqrt{0.5}$ AU).  Also, we find the forced eccentricity ($e_F \approx 0.0257$) using \cite{Quarles2018a} and begin the planetary orbit aligned with the binary in terms of the forced eccentricity, so that the determined obliquity variations are due to changes in the planetary angular momentum vector $L_p$ through the planetary inclination $i_p$ or changes in the spin vector $S_p$.  The planet begins nodally aligned ($\Omega_p=\Omega_{bin}=0^\circ$) with the binary orbit, but inclined by $i_p=10^\circ$.

\begin{figure}
    \centering
    \includegraphics[width=\linewidth]{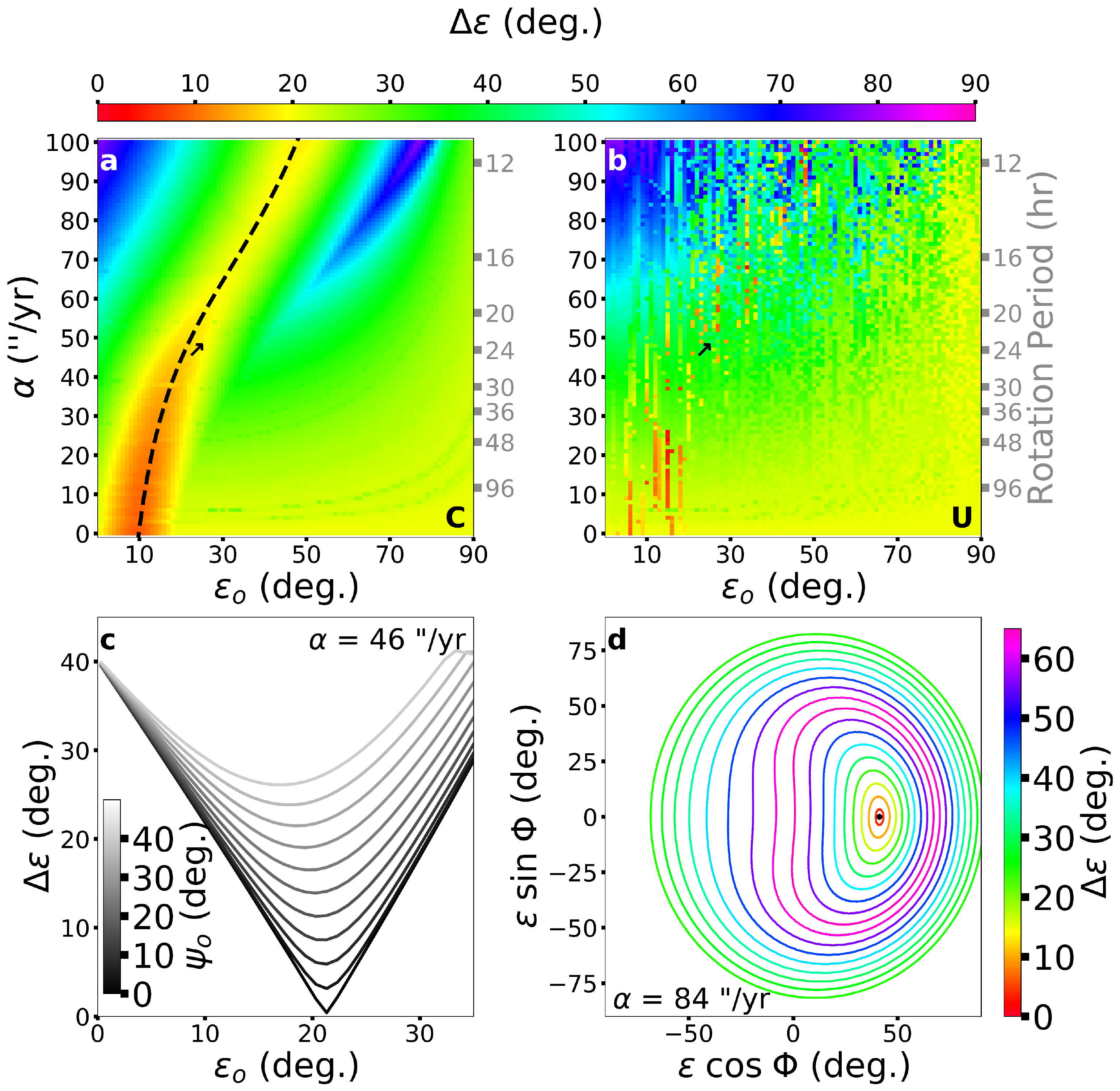}
    \caption{Numerical simulations of the obliquity evolution for a single planet orbiting $\alpha$ Centauri B on an inclined orbit ($i_p=10^\circ$, $\Omega_p=0^\circ$).  Panel (a) illustrates the obliquity variation ($\Delta \epsilon$; color-coded on top) attained as a function of the precession constant $\alpha$ and the initial obliquity $\epsilon_o$ with an identical initial spin longitude value ($\psi_o = 23.76^\circ$ for each simulation). The dashed line marks the initial parameters that minimize $\Delta \epsilon$ when the orbital $L_p$ and spin $S_p$ angular momentum vectors precess together.  Panel (b) shows the outcomes when the initial spin longitude is instead randomly chosen from $0^\circ-360^\circ$.  The arrows in panels (a) and (b) denote the location of a moonless Earth analog within the parameter space.  Panel (c) identifies the changes in location of a fixed point (minimal $\Delta \epsilon$, grayscale) for $\alpha=46$\arcsec/yr when the initial spin longitude $\psi_o$ is nearly aligned and exactly aligned ($\psi_o=\Omega_p$) with the planetary orbit.  Panel (d) displays the phase trajectories for a more rapid spin precession ($\alpha=84$\arcsec/yr), where $\Phi$ ($= \Omega_p - \psi$) represents the canonical variable in the simplified Hamiltonian \citep{Li2014}. }
    \label{fig:single_aCen}
\end{figure}

Figure \ref{fig:single_aCen} shows the results of these simulations (see Sections \ref{sec:nbody} \& \ref{sec:spin}), where each panel (Figs. \ref{fig:single_aCen}a -- \ref{fig:single_aCen}d) highlights various aspects of the dynamics.  One set of simulations, shown in Fig. \ref{fig:single_aCen}a, use a single value for the spin longitude ($\psi_o\approx 23.7^\circ$, \textbf{C}) with each simulation and hence the contours of obliquity variation $\Delta \epsilon$ (color-coded) largely appear smooth.  From solar system and exoplanet studies with single stars \citep{Barnes2016,Shan2018}, we expect a contour ($\alpha \propto 1/\cos \epsilon_o$) to arise that indicates the location of a spin-orbit resonance, where this occurs for slow spin precessions ($\alpha \lesssim 20$\arcsec/yr) in the 30\degree{} variations (light green) nestled within $\sim$20\degree{} variations (yellow-green) from orbital precession ($\Delta \epsilon \approx 2 i_p$).

From our simulations in Fig. \ref{fig:single_aCen}a, there are two regimes with high spin precession constant ($\alpha>70$\arcsec/yr) that allow for $\Delta \epsilon\sim70-80$\degree{} for low ($\epsilon_o<$10\degree) and high ($\epsilon_o\sim$70\degree) initial obliquity.  However, a valley of low obliquity variation ($<$20\degree) appears due to strong spin-orbit coupling, where the spin vector $S_p$ precesses at nearly the same rate as the angular momentum vector $L_p$ does around the normal vector to the binary plane.  Equation \ref{eqn:alpha_prec} relates the precession constant $\alpha$ to the initial obliquity and the dashed line in Fig. \ref{fig:single_aCen}a marks the minimum obliquity variation when the two precession rates ($\dot{\psi}$ \& $g_s$) are commensurate for $i_p = 10^\circ$.  Values of $\Delta \epsilon$ do not reach 0$^\circ$ in Fig. \ref{fig:single_aCen}a because of the initial offset between the spin longitude and the planetary longitude of ascending node (i.e., $\psi_o \neq \Omega_p$).  For slower spin precession constant near the curve (dashed line), lower obliquity variation is possible because higher spin precession constant $\alpha$ has the potential to overlap with the secular precession frequency $g_s$ due to the stellar perturber.  We note the right vertical axis in Fig. \ref{fig:single_aCen}a marks the rotation periods for a planet with a similar composition to the Earth \citep[][see Appendix]{Lissauer2012} and a black arrow denotes the conditions for an Earth-like rotator with modern values ($P_{rot}=23.934$ hr \& $\epsilon_o=23.4^\circ$).

We perform another set of simulations in Fig. \ref{fig:single_aCen}b, with the initial spin longitude $\psi$ chosen randomly for each initial spin state ($\epsilon_o$, $\alpha$) from $0^\circ-360^\circ$ using a uniform distribution (\textbf{U}).  As a result, the contours are not smooth and the obliquity variation is typically $\sim$25$^\circ-35^\circ$ without well-defined regimes of low or high $\Delta \epsilon$.  Since the major difference between Figs. \ref{fig:single_aCen}a \& \ref{fig:single_aCen}b is sampling different values of $\psi_o$, we produce Figs. \ref{fig:single_aCen}c \& \ref{fig:single_aCen}d to better understand these differences.  

In Fig. \ref{fig:single_aCen}c, we perform a small set of simulations for a single $\alpha$ (=46\arcsec/yr).  These simulations vary $\psi_o$ in 4\degree{} steps and $\epsilon_o$ in 1\degree{} steps, where both range from 0$^\circ-40^\circ$.  From the results, we plot the obliquity variation (grayscale) as a function of initial obliquity in order to identify how the point of minimum variation for a given $\psi_o$ changes.  We recover the fixed point from Equation \ref{eqn:alpha_prec} when $\psi_o = \Omega_p = 0^\circ$ and also find that when $\psi_o \gtrsim 30^\circ$, then the minimum obliquity variation increases beyond the expected value from orbital precession ($\Delta \epsilon \approx 2i_p = 20^\circ$).  The non-uniformities and strips found in Fig. \ref{fig:single_aCen}b are due to the $\sim$8\% chance of starting with an appropriate value for $\psi_o$, where $S_p$ \& $L_p$ precess together at nearly the same rate.

Another contributor to the obliquity variation is shown through a resonance diagram using the canonical variables $\epsilon \cos \Phi$ and $\epsilon \sin \Phi$ in Fig. \ref{fig:single_aCen}d, where $\Phi = \Omega_p - \psi$.  For this diagram, we use a faster spin precession constant that is closer to commensurability with $g_s$ to highlight the regime of the largest obliquity variations in Fig. \ref{fig:single_aCen}a from a spin-orbit resonance.  The fixed point in Fig. \ref{fig:single_aCen}d corresponds to $\epsilon_o$ in Fig. \ref{fig:single_aCen}a that intersects the dashed curve and $\alpha = 84$\arcsec/yr.  For initial obliquities near the fixed point, the trajectories librate and produce relatively minor obliquity variations.  This continues until a transition is reached and the trajectories begin to circulate instead of librate.  In the transition region ($\epsilon_o$ $\sim$65\degree), the planet enters a 2:1 resonance between the spin and orbit precession rates that forces much larger obliquity variations. 

\subsection{\emph{Multiple} Planets and a Moon Orbiting $\alpha$ Centauri A or B} \label{sec:multi_results}
In the solar system, secular perturbations (primarily from Jupiter) impact the obliquity evolution and stability of the inner terrestrial planets \citep{Laskar1993c,Innanen1998}.  We evaluate the obliquity evolution for an Earth-like planet initially orbiting in the plane of the binary at inner edge of the conservative habitable zones of either star in $\alpha$ Cen AB, where the planet has terrestrial neighbors that interact dynamically (see Table \ref{tab:SS_analogs}).  The obliquity variation $\Delta \epsilon$ ($=\epsilon_{max} - \epsilon_{min}$) is identified using the spin precession constant $\alpha$ and the initial obliquity $\epsilon_o$ as input parameters, where the initial spin longitude is selected randomly from $0^\circ-360^\circ$ using a uniform distribution (see Section \ref{sec:spin}).  Similar studies were performed using the 8 solar system planets \citep{Laskar1993c}, where we reproduce those results using our methodology.

\subsubsection{Precession due to Neighboring Planets} \label{sec:prec_multi}
\begin{figure}
    \centering
    \includegraphics[width=\linewidth]{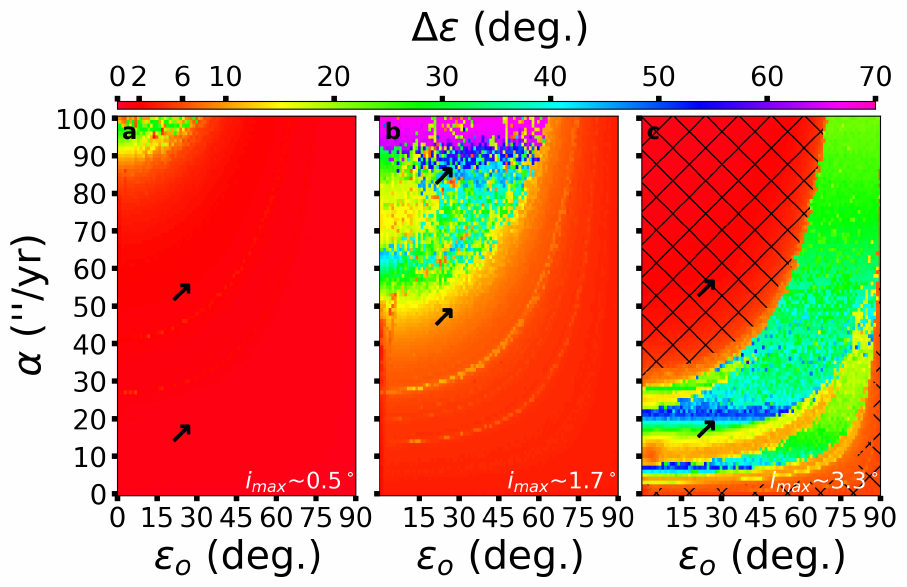}
    \caption{Obliquity variations (color-coded) considering a large range in initial obliquity $\epsilon_o$ and precession constant $\alpha$ for an Earth-like planet orbiting (a) $\alpha$ Centauri A, (b) $\alpha$ Centauri B, or (c) the Sun, for comparison.  The planet is accompanied by 3 terrestrial planet analogs in panels (a) \& (b) (see Section \ref{sec:meth_mult}), where in Panel (c) all 8 planets of the solar system are included.  The black arrows mark the precession constant for a moonless Earth (lower arrow) and an Earth-like planet with a moon similar to our own in terms of mass and orbital distance (i.e., Luna-like).  The hatched regions denote when particularly stable obliquities ($\Delta \epsilon \leq 2i_{max}$) are possible. }
    \label{fig:multis}
\end{figure}

Using the n-body simulations (with 100 year outputs) of the systems with 4 terrestrial planets, we find that the maximum inclination $i_{max}$ achieved for the Earth-like planet orbiting $\alpha$ Cen A, $\alpha$ Cen B, or the Sun is about 0.5\degree, 1.7\degree, and 3.3\degree, respectively.  The maximum inclination calibrates the expectation for the obliquity variation, where the orbital angular momentum vector $L_p$ precesses through one cycle while the spin vector $S_p$ roughly remains fixed (i.e., $\Delta \epsilon \approx 2i_{max}$).  Figure \ref{fig:multis} shows the results of our simulations for each case and marks with arrows 2 spin precession constants with Earth-like initial obliquity ($\epsilon_o=23.4^\circ$): the bottom arrow represents a 23.934 hr rotator and the top arrow denotes the total spin precession constant after adding the induced spin precession from a moon similar in mass and semimajor axis to our own (i.e., Luna) on an Earth-like planet. 
The results in Fig. \ref{fig:multis}a illustrate that a overwhelming majority of initial conditions do not significantly alter the obliquity evolution of the planet.  The median obliquity variation is 1.43\degree{} for our scenario with $\alpha$ Cen A as the host star, where the minimum variation is 1.07\degree, just above the expectation from orbital precession.  Using a Frequency Modified Fourier Transform (FMFT; \cite{Sidlichovsky1996}), the 2 dominant frequencies ($f_1$ \& $f_2$) from orbital perturbations are near $-150$\arcsec/yr and $-97$\arcsec/yr, respectively.  The orbital precession frequency $g_s$ through secular interactions with $\alpha$ Cen B is $\sim$146\arcsec/yr, which is near 1:1 and 3:2 commensurabilities with $f_1$ and $f_2$.  Thus, the small region of higher obliquity variation for low initial obliquity and relatively high spin precession constant ($\sim$96\arcsec/yr) is mainly due to a spin orbit resonance with the stellar binary.  We note that such a large spin precession constant corresponds to a rotation period of $\sim$2--3 hr and it is unphysical to reach overlap with the dominant secular frequency $f_1$.  Although a large moon (i.e., Luna-like) could add to the spin precession of a rapidly rotating Earth-like planet to overlap with $f_2$.  The spin precession constant on the planet due to the host star ($\alpha$ Cen A) is roughly an order-of-magnitude slower and results in a spin that is strongly coupled with the orbit (i.e., $S_p$ and $L_p$ precess together at nearly constant angular separation).  The orbital perturbations from the nearby planets are much weaker than those from the binary companion, which means the timescale for these perturbations to significantly manifest in the obliquity variation are much longer ($\gg$50 Myr) than those considered in this work.

When simulating a similar planetary system orbiting $\alpha$ Cen B (Fig. \ref{fig:multis}b), the obliquity variations are much more varied.  We note that the orbital period of the Earth-like planet is shorter due to the lower luminosity of $\alpha$ Centuari B and our scaling for the conservative habitable zone (see Table \ref{tab:SS_analogs}).  The spin precession constant $\alpha$ is inversely proportional to the planet orbital period and we see the black arrows shifted to higher values of $\alpha$.  The median obliquity variation is 6.00\degree{} for the scenario with $\alpha$ Cen B as the host star, where the minimum variation is 3.53\degree, just above the expectation from orbital precession.  There is a faint contour ($\alpha \propto 1/\cos \epsilon_o$) beginning at $\sim$29\arcsec/yr that is due to a spin-orbit resonance with one of the neighboring planets \citep{Quarles2017,Shan2018}.  The top-left corner displays the most obliquity variation likely due to secular perturbations from both the neighboring planets and the stellar companion, $\alpha$ Cen A.  From the calculation of the secular orbital frequency $g_s$, spin-orbit resonances with the binary orbit that lead to higher obliquity variation are expected when $\alpha \approx 80$\arcsec/yr.  However, the combined coupling between the terrestrial planets and the binary shift allowing for 2 dominant secular modes.  Using a FMFT, the 2 dominant frequencies ($f_1$ \& $f_2$) from orbital perturbations are near $-92$\arcsec/yr and $-58$\arcsec/yr, respectively.  In contrast to Fig. \ref{fig:multis}a, the added spin precession from a Luna-like moon transports the Earth-like planet into a regime of significantly higher obliquity variation ($\Delta \epsilon \sim 30^\circ-50^\circ$).  A faster spinning ($P_{rot}=12$ hr) moonless Earth-like planet could experience obliquity variations near 67\degree, where the range of obliquity extends from $0-67^\circ$.

We reproduce the expected obliquity variations under similar conditions for the Earth within the solar system in Fig. \ref{fig:multis}c.  There are initial conditions (hatched) that act to stabilize the obliquity (i.e., $\Delta \epsilon < 2 i_{max}$) of the Earth typically by an added precession and more specifically due to the presence of our moon (Luna; \cite{Laskar1993a,Laskar1993b,Li2014}).  The strong spin-orbit coupling when $\alpha \gtrsim 50$ \arcsec/yr and $\epsilon_o \lesssim 60^\circ$ leads to low obliquity variations of $\sim 1-2^\circ$, which is below $2i_{max}$ due to nodal precession alone and much smaller than the median obliquity variation (8.26\degree) for our scenario with the solar system.  \cite{Laskar1993c} showed that the overlap of modes induced from secular perturbations between the planets allow for increased obliquity variation and even chaotic obliquity evolution (i.e., large, abrupt changes to obliquity).  By comparison, the zones of obliquity variation in Fig. \ref{fig:multis}b are similar to the 'chaotic sea' in Fig. \ref{fig:multis}c, where we can expect a similar form of chaos.  Earth-like planets orbiting $\alpha$ Cen A (Fig. \ref{fig:multis}a) do not exhibit this phenomenon due to a wide separation of the secular modes.  Another pathway to avoid chaotic or large obliquity variations is through retrograde rotation, which strongly couples the spin and orbit.  We perform simulations of the retrograde-spinning regime ($90^\circ < \epsilon_o < 180^\circ$ ) using the scenarios in Fig. \ref{fig:multis} and find that the obliquity variation for the Earth-like planet is less than or nearly equal to $2i_{max}$ for the full range of spin precession constant ($\alpha \leq 100$\arcsec/yr).  As long as $i_{max}$ remains low, Earth-like planets in retrograde rotation will have stable obliquity.

\begin{figure}
    \centering
    \includegraphics[width=\linewidth]{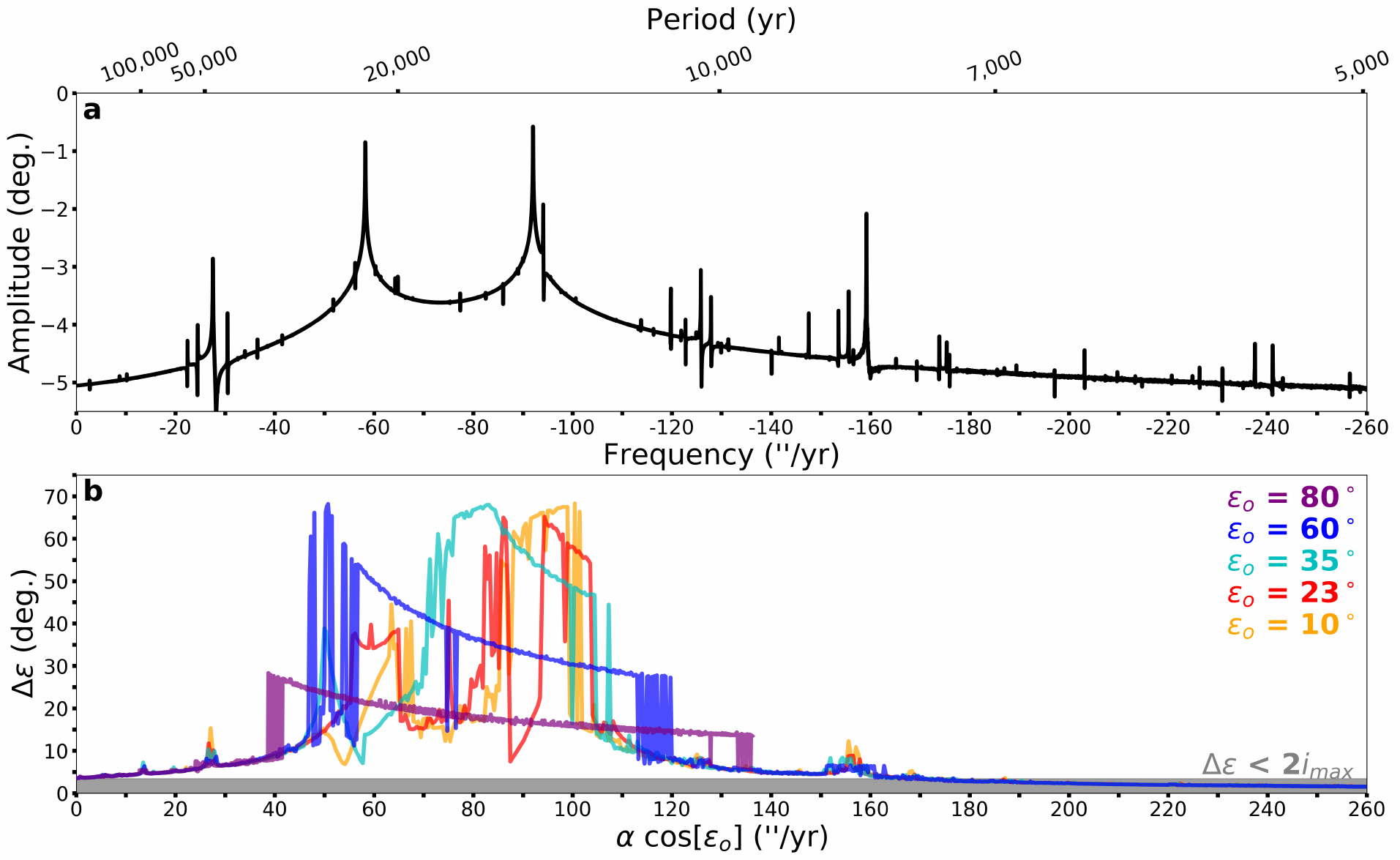}
    \caption{Analysis of an Earth-like planet with 3 terrestrial neighbors orbiting $\alpha$ Centauri B (see Section \ref{sec:meth_mult}) through the (a) Fourier spectrum  of the inclination vector ($i_p e^{i\Omega_p}$) and (b) obliquity variation.  The amplitude from the FFT in panel (a) is given on a base-10 logarithmic scale and the upper horizontal axis marks the period (in yr) for a given frequency (in \arcsec/yr). Panel (b) shows the obliquity variation $\Delta \epsilon$ in terms of the initial spin precession frequency ($\alpha \cos[\epsilon_o]$) for five initial obliquities (color-coded).  The gray bar indicates the region where the obliquity variations are less than those expected from orbital precession ($\Delta \epsilon < 2i_{max}$; see Figure \ref{fig:multis}b).  }
    \label{fig:FFT_aCenB}
\end{figure}

The dominant secular modes for an Earth-like planet orbiting $\alpha$ Cen A are quite high and additional variation may occur for larger $\alpha$ values, but there are physical limits (e.g., possible spin-induce breakup or the existence of a large moon).  As a result, we examine (in Figure \ref{fig:FFT_aCenB}) the Fourier spectrum of the inclination vector ($i_pe^{i\Omega_p}$) of the Earth-like planet orbiting $\alpha$ Cen B and compare with the initial spin precession frequencies ($\alpha\;\cos[\epsilon_o]$ associated with obliquity variation.  Figure \ref{fig:FFT_aCenB}a shows the spectrum of frequencies and corresponding amplitudes (on a base-10 logarithmic scale) using the numerical FFT package within \texttt{sciPy} \citep{Jones2001}.  The upper horizontal axis identifies the corresponding precession timescale (in yr) for a given frequency, where the dominant secular mode forces the planetary inclination vector causing precession on a $\sim$15,000 yr timescale.  Alternatively, a second mode causes precession on a slightly longer $\sim$22,000 yr timescale.  The frequencies associated with these modes are near $-92$\arcsec/yr and $-58$\arcsec/yr, respectively.  Three other modes exist ($\sim$ $-28$, $-128$, and $-159$\arcsec/yr) with lower amplitudes, which lead to smaller obliquity variations. 

We show the degree by which each secular mode affects the obliquity variation over a range of initial obliquities (color-coded) in Figure \ref{fig:FFT_aCenB}b.  The most obliquity variation occurs between the two dominant secular modes and for $\epsilon_o \leq 60^\circ$, the obliquity variation can reach $\sim$65\degree.  Additional excitation of $\Delta \epsilon$ occurs for the other three modes, but with a much smaller amplitude.  Above a spin precession frequency of $\sim$160\arcsec/yr, the spin precession is about a factor of two faster than orbital precession so that the spin $S_p$ and orbital $L_p$ angular momentum vectors effectively move together around the Laplace plane (i.e., normal vector to the binary plane).  Returning to Figure \ref{fig:multis}, we can deduce that Figs. \ref{fig:multis}a \& \ref{fig:multis}b would resemble Fig. \ref{fig:multis}c at higher values of $\alpha$, provided that such ranges are physically accessible.  Figure \ref{fig:multis}b has a zone of stable obliquity, but a spin precession frequency of $\sim$160\arcsec/yr is required, which can be attained through the combination of shorter a spin period ($\sim$6--10 hr), a Luna-like moon, or an appropriate initial obliquity that avoids the secular modes.

\subsubsection{Precession due to Neighboring Planets Including a Moon} \label{sec:prec_moon}
\begin{figure}
    \centering
    \includegraphics[width=\linewidth]{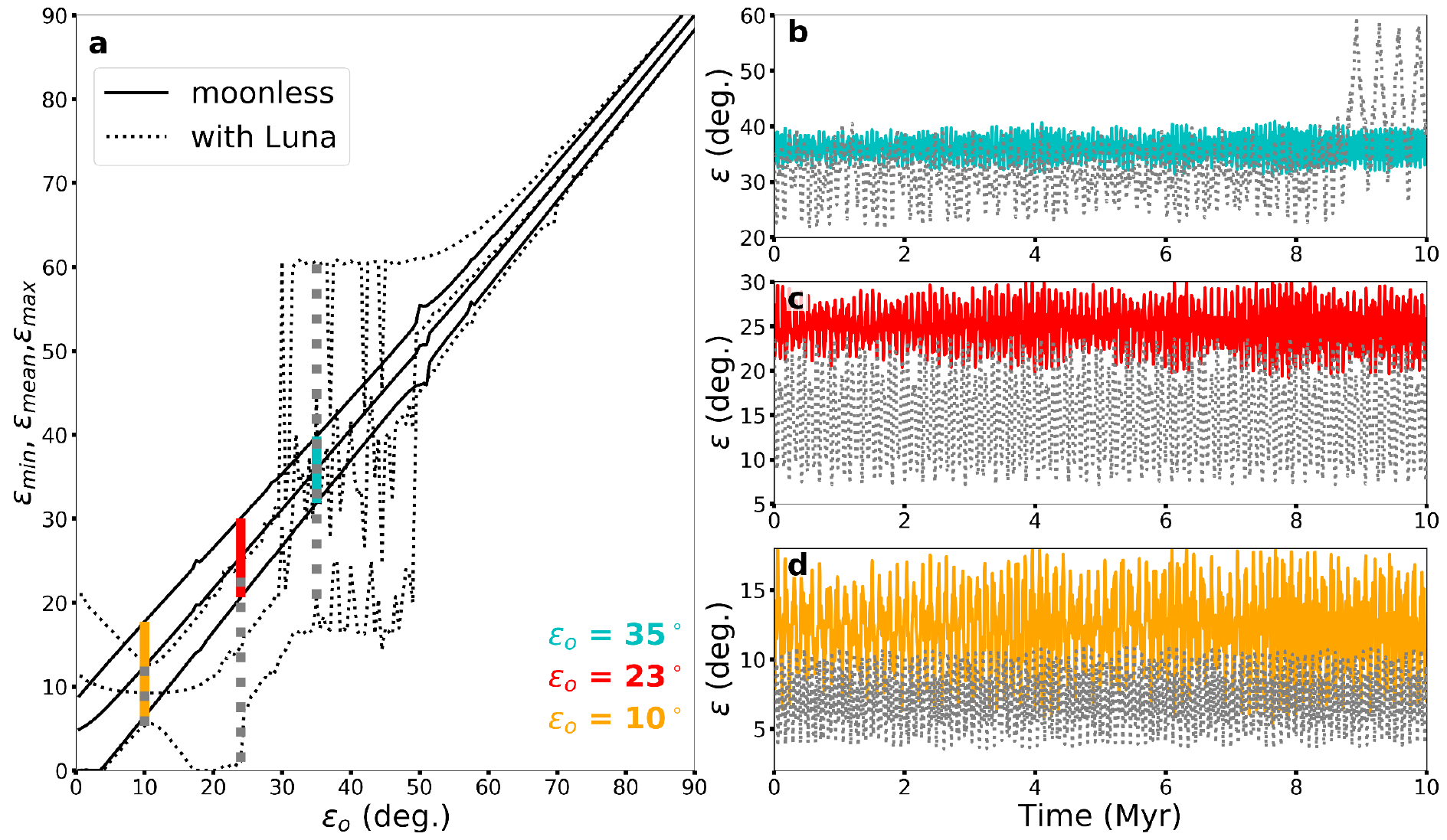}
    \caption{The minimum ($\epsilon_{min}$), mean ($\epsilon_{mean}$), and maximum ($\epsilon_{max}$) obliquity using results from Fig. \ref{fig:multis}b (over 50 Myr) for a precession constant for a moonless Earth-like planet (solid) and an Earth-like planet with a Luna-like moon (dotted).  The range of variation for three obliquities are indicated by vertical lines (gray \& colored).  Panels (b), (c), \& (d) show the time series of the planet's obliquity for the first 10 Myr of each simulation following the line style and color-code from panel (a).}
    \label{fig:moon_aCenB}
\end{figure}

As Figure \ref{fig:multis}c shows, the additional spin precession from a Luna-like moon can transport an Earth-like planet from a state of high or chaotic obliquity variation to a {much smaller} obliquity variation (i.e., stabilized obliquity \cite{Laskar1993a,Laskar1993b,Li2014}).  Figure \ref{fig:multis}b displays a similar transport, but the end state is one of high or chaotic obliquity.  In Figure \ref{fig:moon_aCenB}a, we show the minimum, mean, and maximum obliquity attained over a range of initial obliquities ($\epsilon_o \leq 90^\circ$) for two precession constants (46\arcsec/yr \& 84\arcsec/yr) that correspond to an Earth-like planet ($P_{rot}=23.934$ hr) with (dashed) and without (solid) a moon like our own (i.e., Luna-like).  Three initial obliquities (color-coded) are highlighted in Fig. \ref{fig:moon_aCenB}a, where the corresponding time series for the first 10 Myr are given in Figs. \ref{fig:moon_aCenB}b--\ref{fig:moon_aCenB}d.  Fig. \ref{fig:moon_aCenB}a shows a chaotic bridge from $\sim$ $30^\circ \leq \epsilon_o \leq 50^\circ$ when a Luna-like moon is included, where the obliquity variation ($\epsilon_{max} - \epsilon_{min}$) without such a moon is more regular.  Chaotic variations still exist within the obliquity time series without a moon (colored lines; Figs. \ref{fig:moon_aCenB}b--\ref{fig:moon_aCenB}d), but the overall obliquity variation is limited to $\sim$10\degree and this variation is larger than the 3.4\degree{} expected from orbital precession alone.  The chaotic bridge is largely due to the second secular mode ($f_2 \sim 59$\arcsec/yr), where the initial spin precession frequency ($\alpha\;\cos[\epsilon_o]$) overlaps with this range of initial obliquities and produces up to $\sim$40\degree{} obliquity variations after 50 Myr of integration time.  In Fig. \ref{fig:moon_aCenB}b a transition occurs after $\sim$9 Myr, where the obliquity variation approximately doubles about a different mean obliquity altogether, while the other initial obliquities (23\degree{} and 10\degree) with a Luna-like moon outside the chaotic bridge (Figs. \ref{fig:moon_aCenB}c and \ref{fig:moon_aCenB}d, respectively) oscillate with a mean obliquity near the initial value.  At large initial obliquities ($\epsilon_o \gtrsim 65^\circ$), the obliquity variation for both cases (with and without a Luna-like moon) in Fig. \ref{fig:moon_aCenB}a are similar and narrower because the initial spin precession frequency drops below 40\arcsec/yr, where the spin precession is now much slower than the orbital precession.  Overall, a Luna-like moon greatly increases the obliquity variation for initial obliquities less than 60\degree, while the absence of such a moon allows for a lower bounded variation.

\begin{figure}
    \centering
    \includegraphics[width=\linewidth]{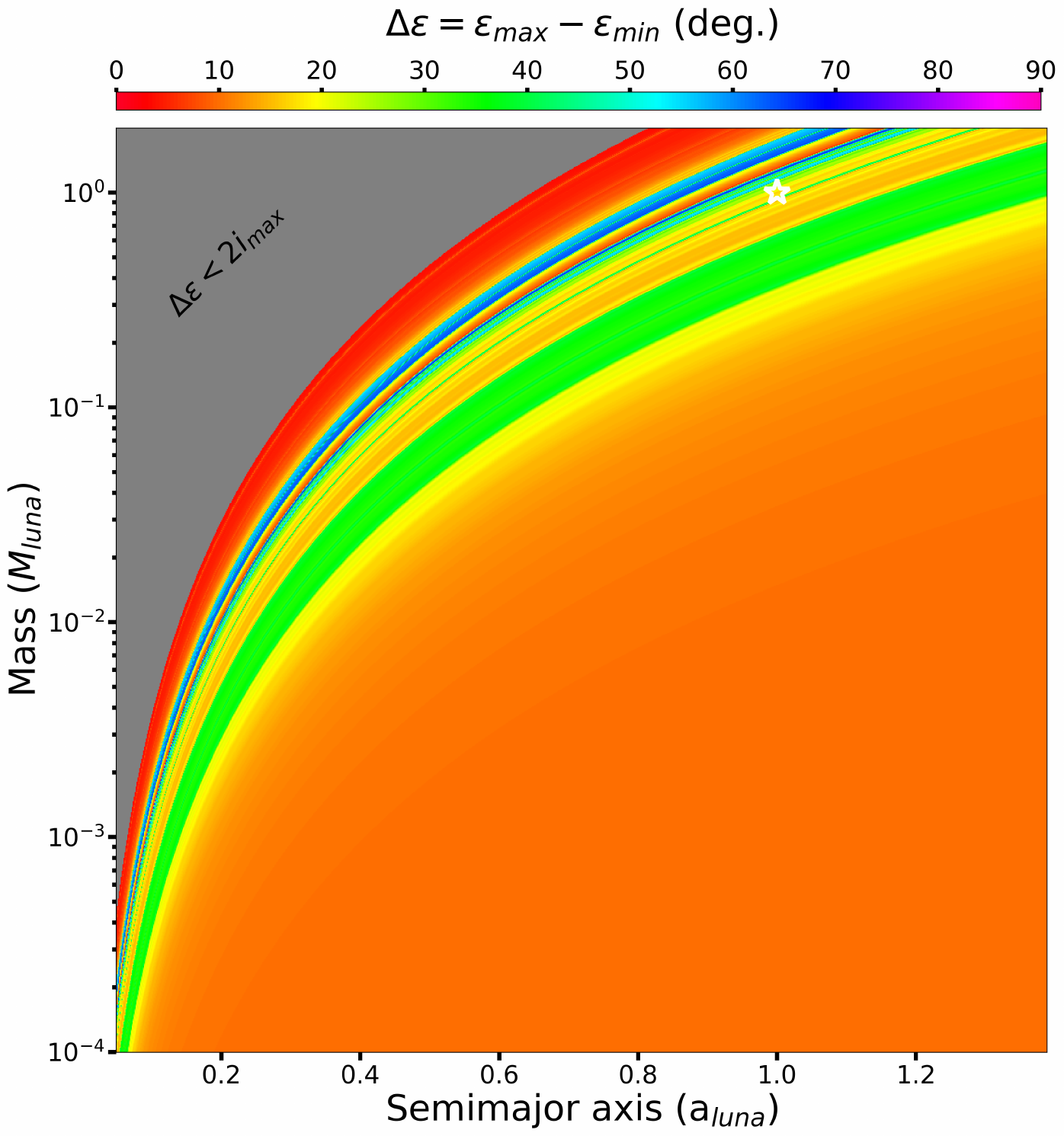}
    \caption{Obliquity variations (color-coded) for an Earth-like rotator ($P_{rot}=23.934$ hr \& $\epsilon_o = 23.4^\circ$) with 3 terrestrial neighbors and a moon orbiting $\alpha$ Centauri B (see Section \ref{sec:meth_mult}).  The lunar parameters (mass \& semimajor axis) are used to increase the initial value for the spin precession constant $\alpha$ and the units are given in terms of our moon (Luna; see Section \ref{sec:meth_mult}).  The gray region denotes the parameters where the spin vector $S_p$ precesses much faster than the orbital angular momentum vector $L_p$ and the obliquity variation is minimized.  The orange region that covers most of the bottom right corresponds to an obliquity variation ($\sim$17$^\circ$) where the added precession due to a moon is negligible.  The white star marks the obliquity variation for an Earth-like planet from a Luna-like moon (see Fig. \ref{fig:multis}b). }
    \label{fig:moon_var}
\end{figure}

Figure \ref{fig:multis}b shows that the added spin precession from a Luna-like moon increases the obliquity variation of an Earth-like planet orbiting $\alpha$ Cen B, but a different type of moon, in terms of its mass or semimajor axis, may have a negligible or more beneficial effect.  We explore the possible obliquity variations for a wide range of moon parameters using Equation \ref{eqn:moon_prec} (see Section \ref{sec:meth_mult}) in Figure \ref{fig:moon_var}.  The Earth-like planet (using orbital parameters from Table \ref{tab:SS_analogs}) begins the simulation as an Earth-like rotator ($P_{rot}=23.934$ hr \& $\epsilon_o = 23.4^\circ$) and the spin precession constant is augmented by the assumed mass (in Lunar masses M$_{luna}$) and semimajor axis (in Lunar distances $a_{luna}$).  The white star in Fig. \ref{fig:moon_var} marks the obliquity variation using a precession for a Luna-like moon found in Figs. \ref{fig:multis}b and \ref{fig:moon_aCenB}c.  There is a somewhat narrow range of precession frequencies (see Figure \ref{fig:FFT_aCenB}a) that produces very large obliquity variations without a moon and as a result, there is a narrow portion of this parameter space where a moon would increase the obliquity variation over the $\sim$17\degree{} (orange region in Fig. \ref{fig:moon_var}) when the moon is either to widely separated or to minuscule in mass.  Conversely, there is a regime where the combined spin precession from the Earth-like planet's rotation and that from the moon is faster than the orbital precession induced from the stellar binary (red \& gray region in Fig. \ref{fig:moon_var}).  Massive and close-in moons allow for a very rapid spin precession where the obliquity variation is minimized in a similar manner as in the solar system \citep{Laskar1993b}.  However, tidal interactions between the moon and its Earth-like host can push the moon outward (increasing semimajor axis) so that the epoch of stabilized obliquity variations may be fleeting.  Detailed calculations of such tidal interactions are beyond the scope of this work.

\subsection{Flux Variations for Earth-like Planets Orbiting $\alpha$ Centauri B}
The variation of the Earth's obliquity due to astrophysical forces correlates with historical climate cycles \citep{Kerr1987, Mitrovica1995,Pais1999}, where the foremost example arises from geologic data for ice ages from \cite{Milankovictch1969}.  Recently, \cite{Deitrick2018} numerically reproduced Milankovitch cycles for the Earth and a few exoplanets that orbit single Sun-like stars using the \texttt{VPlanet} software package \citep{Barnes2019}, which relies on a long list of model assumptions.  For our investigation, we focus on calculations of the top of the atmosphere radiative flux (see Section \ref{sec:flux_calc}) because it relies only on a small number of assumptions (e.g., thin \& at least partially transparent atmosphere), which is preferred since planets have yet to be confirmed around either star in $\alpha$ Cen AB; although the search for planets there is ongoing \citep{Belikov2017a,Belikov2017b,Bendek2018,Sirbu2017,Sirbu2018}.  Our calculations are also limited to considering an Earth-like planet orbiting $\alpha$ Cen B under a few scenarios: a single planet ($i_p = 10^\circ$ or $i_p = 2^\circ$), a system of 4 terrestrial planets similar to the solar system, and a system of 4 terrestrial planets where the Earth-like world hosts a Luna-like moon.  Similar calculations could be performed considering scenarios around $\alpha$ Cen A, but the lower obliquity variations ($\Delta \epsilon \lesssim 1.5^\circ$, Figure \ref{fig:multis}a) would produce results largely similar to the Earth with more pronounced flux changes at orbital extremes due to a higher planetary eccentricity ($e_p\sim 0.05$, \cite{Quarles2018a}).

\begin{figure}
    \centering
    \includegraphics[width=\linewidth]{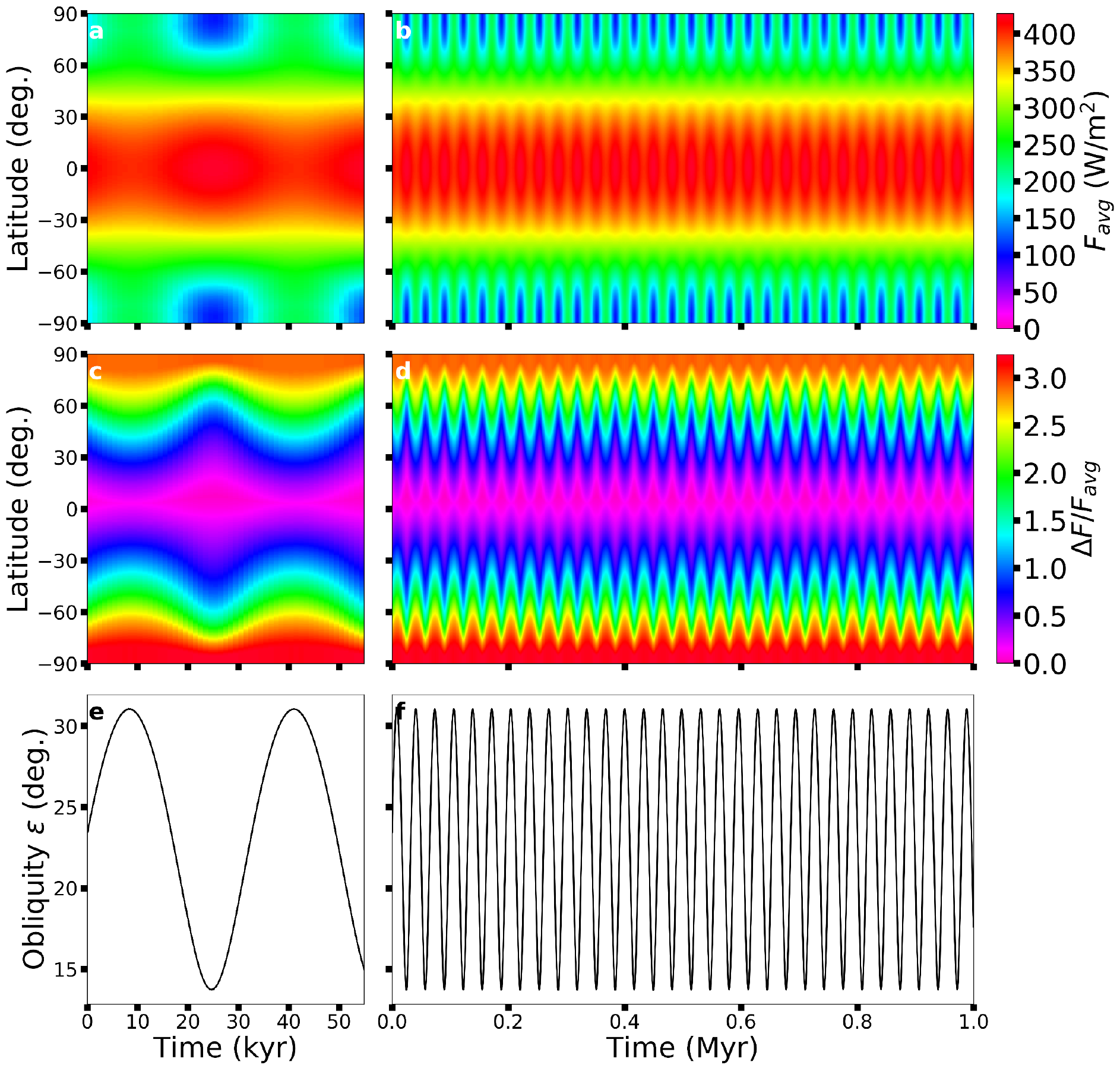}
    \caption{Flux variations (a--d) and obliquity evolution (e \& f) for a \emph{single} inclined Earth-like rotator ($P_{rot}=23.934$ hr, $i_p = 10^\circ$, \& $\epsilon_o = 23.4^\circ$) without a moon orbiting $\alpha$ Centauri B.  The flux variations (a--d) are given in terms of the orbitally averaged flux $F_{avg}$ as a function of latitude and the fractional change in flux $\Delta F/F_{avg}$ per orbit.  The left column (a, c, \& e) highlights the variations over a timescale similar to the modern Earth, while the right column (b, d, \& f) shows the variation over a million years.  }
    \label{fig:single_fluxvar10}
\end{figure}

The flux variation and obliquity evolution of a \emph{single} Earth-like planet orbiting $\alpha$ Cen B inclined by 10\degree{} is shown in Figure \ref{fig:single_fluxvar10}, where the left and right columns show the short and long term variations, respectively.  We calculate the orbitally averaged flux $F_{avg}$ as a function of latitude (Figs. \ref{fig:single_fluxvar10}a \&  \ref{fig:single_fluxvar10}b) to demonstrate that $F_{avg}$ near the equator is similar to Earth-like conditions \citep{Quarles2017}, but the polar regions are quite different.  The average flux at the poles for the modern Earth (with our moon) is between 150--200 W/m$^2$, where obliquity variations correlate with the 50 W/m$^2$ variance on a timescale of $\sim$41,000 years.  Figure \ref{fig:single_fluxvar10}a shows Earth-like conditions can be achieved for only $\sim$1,000 yr intervals within the spin precession cycle.  There is a transition period of $\sim$10,000 yr and then $F_{avg}$ falls to $\lesssim 100$ W/m$^2$.  The fractional change of flux $\Delta F/F_{avg}$ in Figs. \ref{fig:single_fluxvar10}c \& \ref{fig:single_fluxvar10}d probes when the change in flux over an orbit is significantly more or less than the average flux.  During the eras of relatively large $F_{avg}$ at the poles, the southern pole changes by a factor of 3 while the northern pole experiences slightly less variation.  These variations describe a range of possible climatic changes, where ice ages or polar glaciation appear likely when the changes in $F_{avg}$ are dramatic.  Studies of past states of the Martian atmosphere suggest that changes in atmospheric pressure coupled with large changes in planetary obliquity (similar to those present in Fig. \ref{fig:single_fluxvar10}) are responsible for the inhospitable state of the present Martian climate \citep{Head2003,Head2005,Mischna2013,Mansfield2018,Kite2019}.  Figs. \ref{fig:single_fluxvar10}e \& \ref{fig:single_fluxvar10}f provide the short and long term obliquity oscillations, where a minimum in obliquity to nearly 14.5\degree{} drives the shifts in flux and possibly climates.  Due to the small number of external forces, the long term variations, although dramatic, are quite regular.

\begin{figure}
    \centering
    \includegraphics[width=\linewidth]{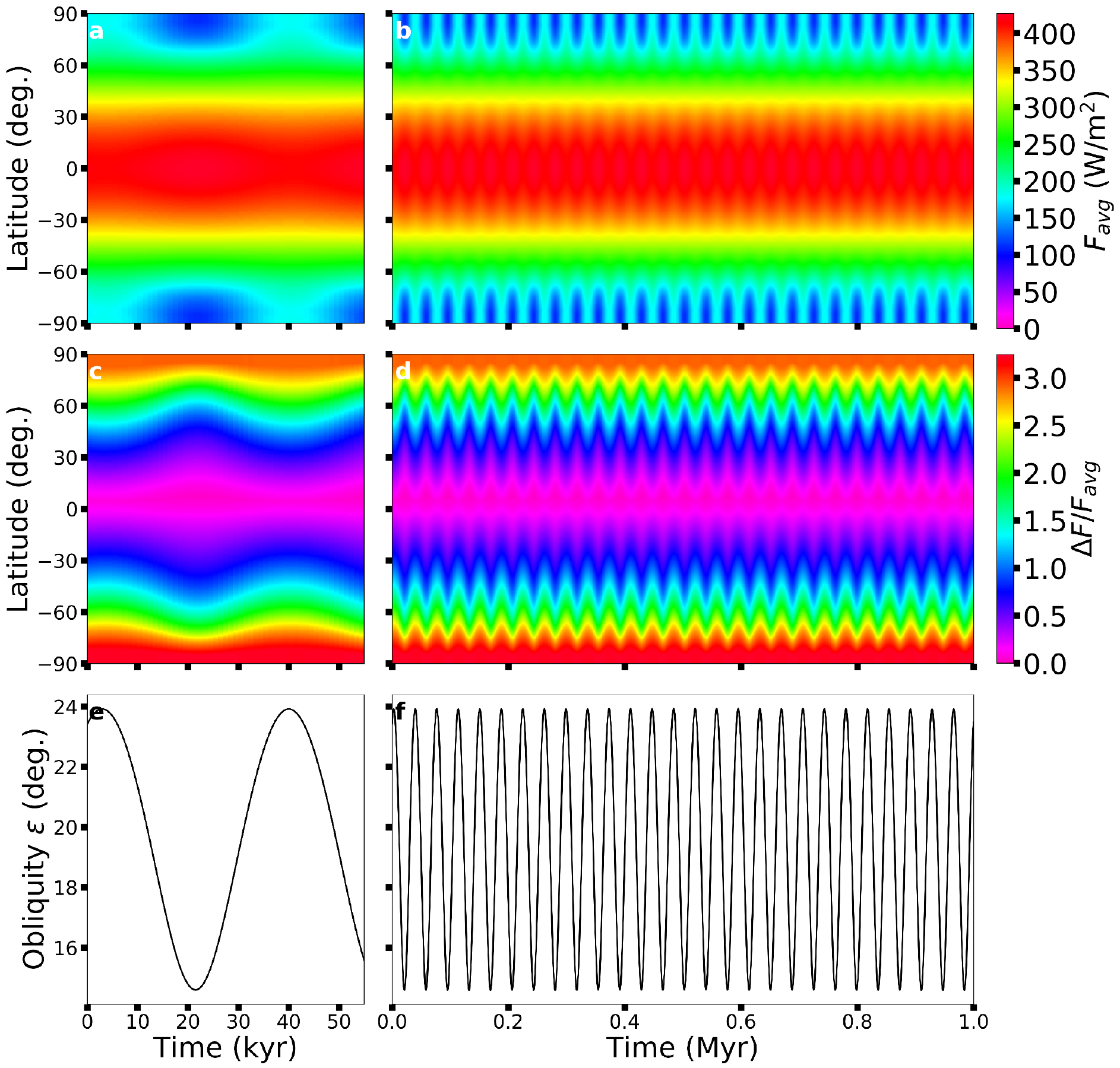}
    \caption{Similar to Fig. \ref{fig:single_fluxvar10}, where the Earth-like rotator ($P_{rot}=23.934$, $i_p = 2^\circ$, \& $\epsilon_o = 23.4^\circ$) is \emph{less inclined}. }
    \label{fig:single_fluxvar2}
\end{figure}

The mutual inclination of disks in stellar binaries are estimated to be $\sim$10\degree, but this is uncertain due to the relatively small number of observations using polarimetry \citep{Monin2006}. We explore the flux variations of a system similar to Fig. \ref{fig:single_fluxvar10}, where the planetary orbit is initially less inclined ($i_p = 2^\circ$) to the binary plane in Figure \ref{fig:single_fluxvar2}.  The orbit-averaged flux $F_{avg}$ in Figs. \ref{fig:single_fluxvar2}a \& \ref{fig:single_fluxvar2}b reaches Earth-like values for about half of the spin precession cycle, while the other half of the cycle exhibits lower flux ($F_{avg}\lesssim 100$ W/m$^2$.  This cycle contributes to large changes at the poles (Figs. \ref{fig:single_fluxvar2}c \& \ref{fig:single_fluxvar2}d), but the changes to extreme $\Delta F/F_{avg}$ are more gradual than for the higher inclination case (Figs. \ref{fig:single_fluxvar10}c \& \ref{fig:single_fluxvar10}d).  The obliquity variation shown in Figs. \ref{fig:single_fluxvar2}e \& \ref{fig:single_fluxvar2}f is $\sim$10\degree{}, which is 2.5$\times$ larger than the expectation from orbital precession alone ($\approx$4\degree).  The long-term variation is regular with oscillations on timescales similar to the Earth.  

\begin{figure}
    \centering
    \includegraphics[width=\linewidth]{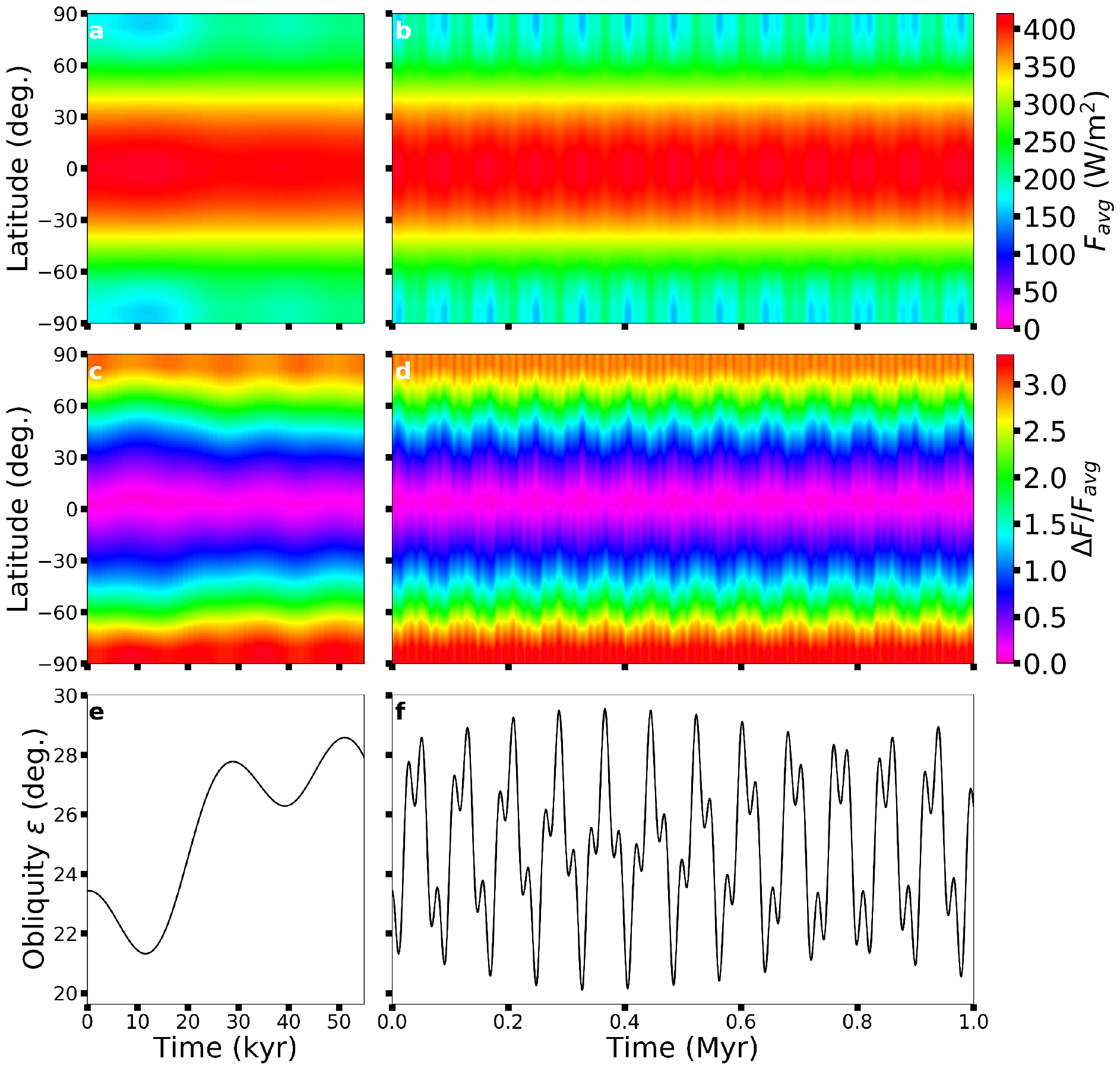}
    \caption{Similar to Fig. \ref{fig:single_fluxvar10}, where the Earth-like rotator ($P_{rot}=23.934$ \& $\epsilon_o = 23.4^\circ$) is \emph{nearly coplanar and accompanied by neighboring terrestrial planet analogs} (see Section \ref{sec:meth_mult}).  The overall variation ($\Delta \epsilon$) is similar in magnitude to Fig \ref{fig:single_fluxvar2}, but the change in precession induced by the neighboring planets limits the oscillations in the short term behavior.  As a result, the flux variations at the poles are modest. }
    \label{fig:multi_fluxvar}
\end{figure}

Figure \ref{fig:multi_fluxvar} demonstrates the importance of nearby terrestrial planets that mitigate the forcing due to the stellar companion (see Section \ref{sec:prec_multi}).  The orbit-averaged flux $F_{avg}$ (Figs. \ref{fig:multi_fluxvar}a \& \ref{fig:multi_fluxvar}b) depicts more Earth-like conditions, where the values at the poles oscillate between 150--200 W/m$^2$ due to the bounds of obliquity variation $\Delta \epsilon$ (Figs. \ref{fig:multi_fluxvar}e \& \ref{fig:multi_fluxvar}f).  The fractional change in the orbit-averaged flux $\Delta F/F_{avg}$ is also more regular (Figs. \ref{fig:multi_fluxvar}c \& \ref{fig:multi_fluxvar}d) over the short and long term, respectively.  Large fluctuations over an orbit still occur at the poles, but this is similar to Earth-like activity.  Figs. \ref{fig:multi_fluxvar}e \& \ref{fig:multi_fluxvar}f show the irregularity of the obliquity over time, where there are rapid variations ($\sim 10,000-20,000$  yr), medium variations ($\sim 100,000$ yr), and slow variations ($\sim 1$ Myr).

\begin{figure}
    \centering
    \includegraphics[width=\linewidth]{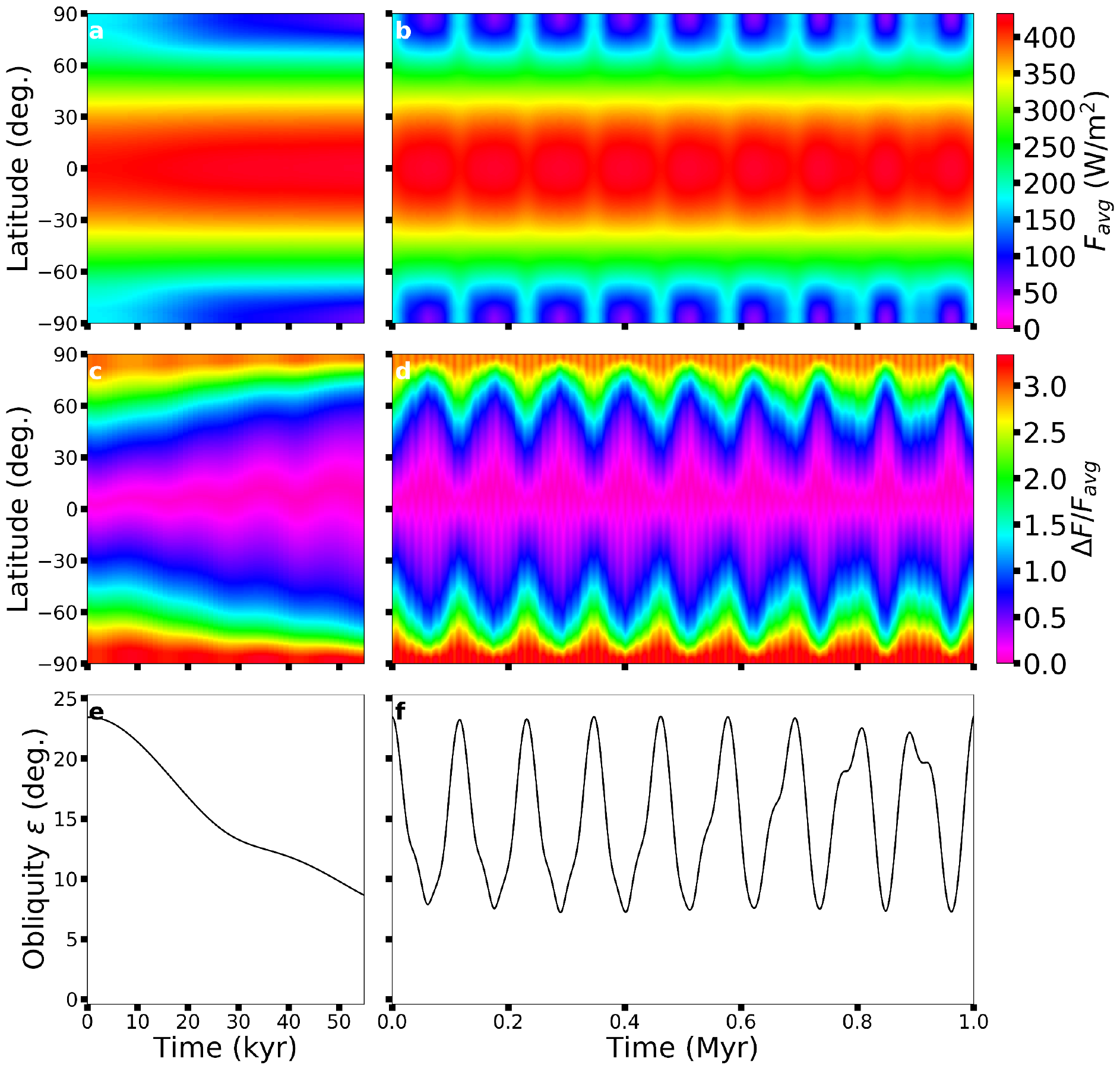}
    \caption{Similar to Fig. \ref{fig:multi_fluxvar}, where the Earth-like rotator ($P_{rot}=23.934$ \& $\epsilon_o = 23.4^\circ$) is nearly coplanar, accompanied by neighboring terrestrial planet analogs, and \emph{has a Luna-like moon} (see Section \ref{sec:meth_mult}). }
    \label{fig:moon_fluxvar}
\end{figure}

One might assume that a large moon similar to our own (i.e., Luna-like) would drive the system to more Earth-like conditions (i.e., further restrict the bounds of obliquity variation \cite{Laskar1993b}), but this is not the case (see Section \ref{sec:prec_moon}) as shown in Figure \ref{fig:moon_fluxvar}.  The orbit-averaged flux (Figs. \ref{fig:moon_fluxvar}a \& \ref{fig:moon_fluxvar}b) at the poles is not as mild the multiple planet case without a moon (Fig. \ref{fig:multi_fluxvar}).  The amplitude of the obliquity variation (Figs. \ref{fig:moon_fluxvar}e \& \ref{fig:moon_fluxvar}f) is not as large as in Fig. \ref{fig:single_fluxvar10}, but the frequency is lower.  This results in the longer duration of extreme lows in $F_{avg}$ at the poles that span nearly 100,000 yr and is punctuated by brief interludes ($\sim$10,000 yr) where Earth-like conditions can return.  Figs. \ref{fig:moon_fluxvar}e \& \ref{fig:moon_fluxvar}f shows slow changes to the obliquity on the 50,000 timescale, where somewhat regular variation occurs over 100,000 yr timescales.  If the initial obliquity begins instead $\sim$35\degree, then this regular variation could be temporary and a larger obliquity variation can occur for future states (see Fig. \ref{fig:moon_aCenB}b).  In such states, the maximum obliquity would be sufficiently high for ice-belts at the equator to develop \citep{Kilic2018} and introduce another set of concerns for potentially {habitable} climates to form.  On the other hand, an initial obliquity that begins $\sim$20\degree would have a larger amplitude oscillation with a minimum obliquity $\epsilon_{min}$ nearly reaching 0\degree{}, but the period of these oscillations would be similar to those in Fig. \ref{fig:moon_fluxvar}f.  In contrast to the solar system, a Luna-like moon in this scenario would destabilize potential climates or at least alter them significantly due to the slow large changes to obliquity.

\subsection{Likelihood of Stable Obliquity for a \emph{Single} Inclined Earth-mass Planet Orbiting a Solar-Type Primary Star}
\label{sec:gen_bin}
Our results for a single planet orbiting $\alpha$ Cen B (see Section \ref{sec:single_aCenB}) show that the planetary mutual inclination $i_p$ (relative to the binary plane) affects the possible obliquity variation $\Delta \epsilon$.  In particular, the nodal precession of the planetary orbit sets a lower bound on the obliquity variation ($\Delta \epsilon \approx 2 i_p$) when the initial spin longitude $\psi_o$ is randomly selected (Fig. \ref{fig:single_aCen}b).  We evaluate here the conditions for obliquity variation that are dependent on the binary orbit by varying the: (1) precession (see \emph{Set 1} in Section \ref{sec:bin_set1}) and (2) the forcing (see \emph{Set 2} in Section \ref{sec:bin_set2}) due to the binary.  Furthermore, we generalize these results using Monte Carlo methods to identify the probability that a single inclined planet orbiting at the inner edge of the primary star's conservative habitable zone would experience a stable obliquity ($\Delta \epsilon \leq 2.4^\circ$) for a given binary semimajor axis $a_{bin}$.

\subsubsection{Obliquity Variations due to Nodal Precession from Star B}
\begin{figure}
    \centering
    \includegraphics[width=\linewidth]{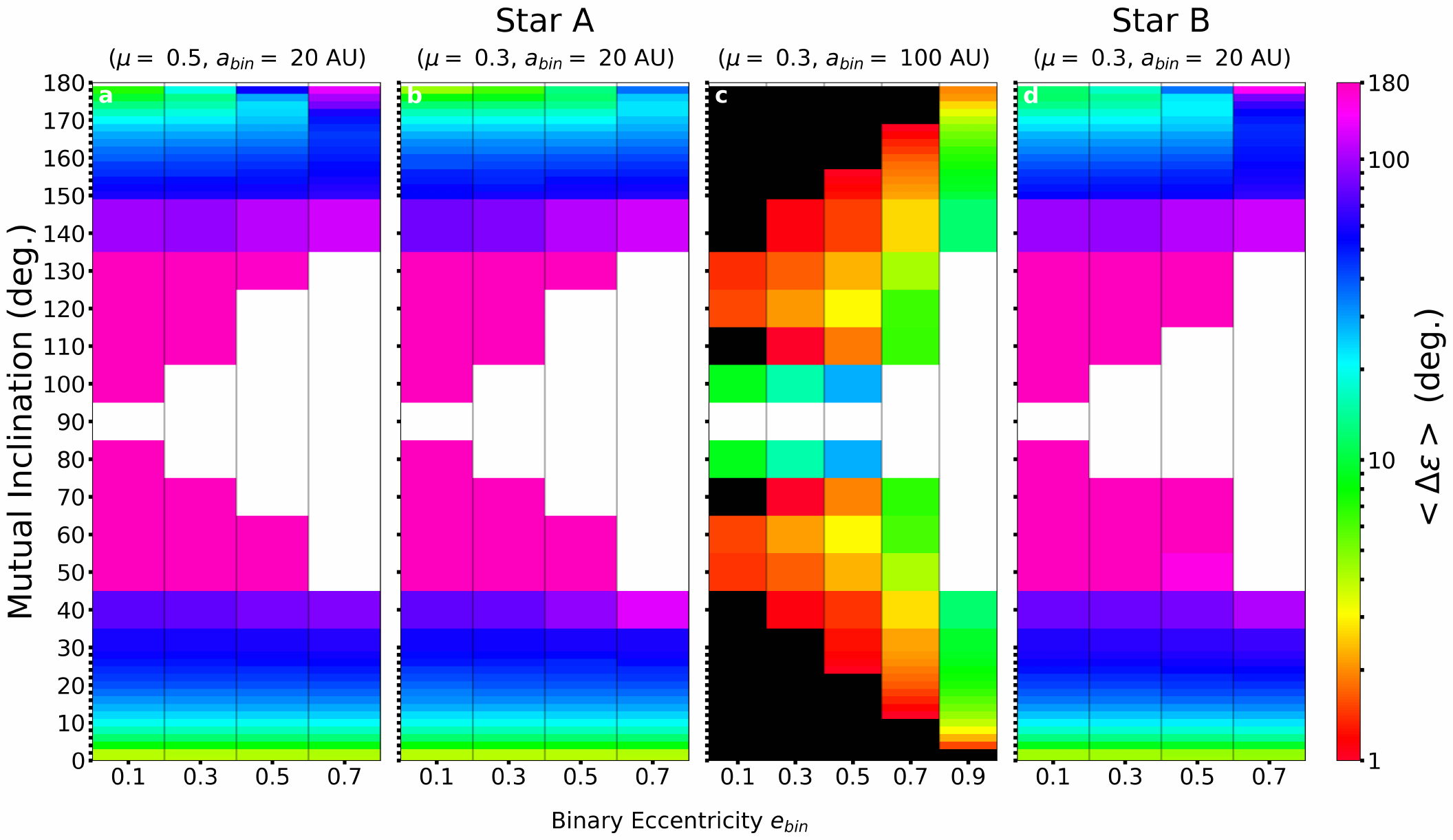}
    \caption{Median obliquity variation $\left<\Delta \epsilon \right>$ (color-coded) for a \emph{single} Earth-like planet around star A (a--c) or star B (d) over a range for planetary mutual inclinations and binary eccentricity $e_{bin}$.  Each cell evaluates a range of spin parameters ($\epsilon_o$ \& $\alpha$) to determine the median value (see Section \ref{sec:gen_bin}) for the planet.  Panels (a) and (b) depict stars that are separated by 20 AU, but the mass ratio between the star varies ((a) $\mu = 0.5$ \& (b) $\mu = 0.3$).  Panel (c) uses the same mass ratio as panel (b) and the binary separation is increased to 100 AU.  Panel (d) shows similar simulations to those in panel (b), except the planet now orbits the other stellar component.  Black cells mark initial conditions where the median obliquity variation is less than 1\degree and white cells signify conditions that are not simulated due to orbital instabilities. }
    \label{fig:single_orb}
\end{figure}

Figure \ref{fig:single_orb} shows the results of our simulations (color-coded on a logarithmic scale) for \emph{Set 1}(see Section \ref{sec:bin_set1}) measuring the median obliquity variation $\left<\Delta \epsilon \right>$, where the binary eccentricity $e_{bin}$ and planetary mutual inclination $i_p$ are varied.  In Fig. \ref{fig:single_orb}a, \ref{fig:single_orb}b, \& \ref{fig:single_orb}d, the stars are closely separated ($a_{bin} = 20$ AU) so that an extreme binary eccentricity (e.g., $e_{bin} = 0.9$) destabilizes the orbits of all planets within the host star's conservative habitable zone.  As a result, we do not evaluate the obliquity variation for $e_{bin} = 0.9$, and note that in the case of Fig. \ref{fig:single_orb}c the wide separation ($a_{bin}=100$ AU) does allow for stable planetary orbits within Star A's conservative habitable zone.  The white cells Fig. \ref{fig:single_orb} mark those conditions where orbital stability is not permitted due to the KL mechanism \citep{Kozai1962,Lidov1962} and the black cells in Fig. \ref{fig:single_orb}c signify that $\left<\Delta \epsilon \right>$ is less than 1\degree.

For the equal-mass star scenario (Fig. \ref{fig:single_orb}a), prograde orbits that are not influenced by the KL mechanism follow our expectation from orbital precession ($\left<\Delta \epsilon\right> = 2 i_p$).  After the initial mutual inclination enters the regime for the KL mechanism ($40^\circ \lesssim i_p \lesssim 140^\circ$), the obliquity variation dramatically increases and allows for retrograde rotation.  The backward rotation (retrograde) of the planet is largely due to how obliquity is defined and that the eccentric KL mechanism \citep{Lithwick2011,Li2014b} flips the orbital angular momentum vector $L_p$ (Fig. \ref{fig:spin_vector}), while the spin angular momentum vector $S_p$ is unchanged relative to the normal vector to the binary orbit $L_{bin}$ (i.e., Laplace plane).  Orbital flips due to the eccentric KL mechanism typically occur in our simulations on million year timescales \citep{Lithwick2011}, where our total integration time is 50 Myr and thus this phenomenon occurs many times to dominate our measure of the median obliquity variation $\left<\Delta \epsilon \right>$.  The same physical processes occur for our simulations of unequal mass binaries (Figs. \ref{fig:single_orb}b \& \ref{fig:single_orb}d).  The retrograde-orbiting region that is beyond the KL mechanism ($i_p \geq 150^\circ$) has lower median obliquity variations, but the variation across the binary eccentricity is not symmetric with the respective prograde-orbiting region.  Although this regime allows for larger separations from the host star for stability \citep{Henon1970}, the forced eccentricity $e_F$ due to the stellar companion scales approximately linearly ($e_F \propto e_{bin}/(1-e_{bin}^2)$; \cite{Heppenheimer1978,Andrade2016,Quarles2018a}) and the $e_F$ for the prograde-orbiting region scales differently \citep{Andrade2016,Andrade2017}.  

In the large $\mu$ and $e_{bin}$ regime, the eccentricity term in the equations of motion (Equation \ref{eqn:eom}) dominates and produces larger obliquity variations. Figs. \ref{fig:single_orb}b and \ref{fig:single_orb}d show a similar feature, where in Fig. \ref{fig:single_orb}b the effect is minimal because the mass ratio $\mu$ is smaller and in Fig. \ref{fig:single_orb}d the effect is enhanced because the planet now orbits a lower mass star (M$_B$ = 0.8 M$_\odot$) with a more massive perturber (M$_A$ = 1.86 M$_\odot$).  Fig. \ref{fig:single_orb}c illustrates the scenario for a planet orbiting Star A, where the forced eccentricity $e_F$ is reduced by a factor of 5 and the orbital precession frequency is reduced by a factor of 5$^3 (= 125)$ compared to Fig. \ref{fig:single_orb}b.  This means that the planetary orbital variations (due to KL mechanism and nodal precession) occur on $1-2$ Myr timescales, where the spin precession timescale is typically an order of magnitude faster.  As a result, the spin angular momentum $S_p$ precesses relatively quickly about $L_p$ while $L_p$ slowly precesses around $L_{bin}$ and most cases produce a median obliquity variation less than 10\degree.  The orbital precession timescale increases to $\sim$10 Myr when the binary eccentricity $e_{bin}$ is reduced to 0.1 and hence the largest opportunity for low median obliquity variations occur for a nearly circular binary.  The KL mechanism scales in a similar fashion with binary eccentricity $e_{bin}$ and that regime ($40^\circ \lesssim i_p \lesssim 140^\circ$) now allows for more orbital stability and relatively high median obliquity variations in Fig. \ref{fig:single_orb}c.

\subsubsection{Obliquity Variations due to Spin-orbit Resonances}
\begin{figure}
    \centering
    \includegraphics[width=\linewidth]{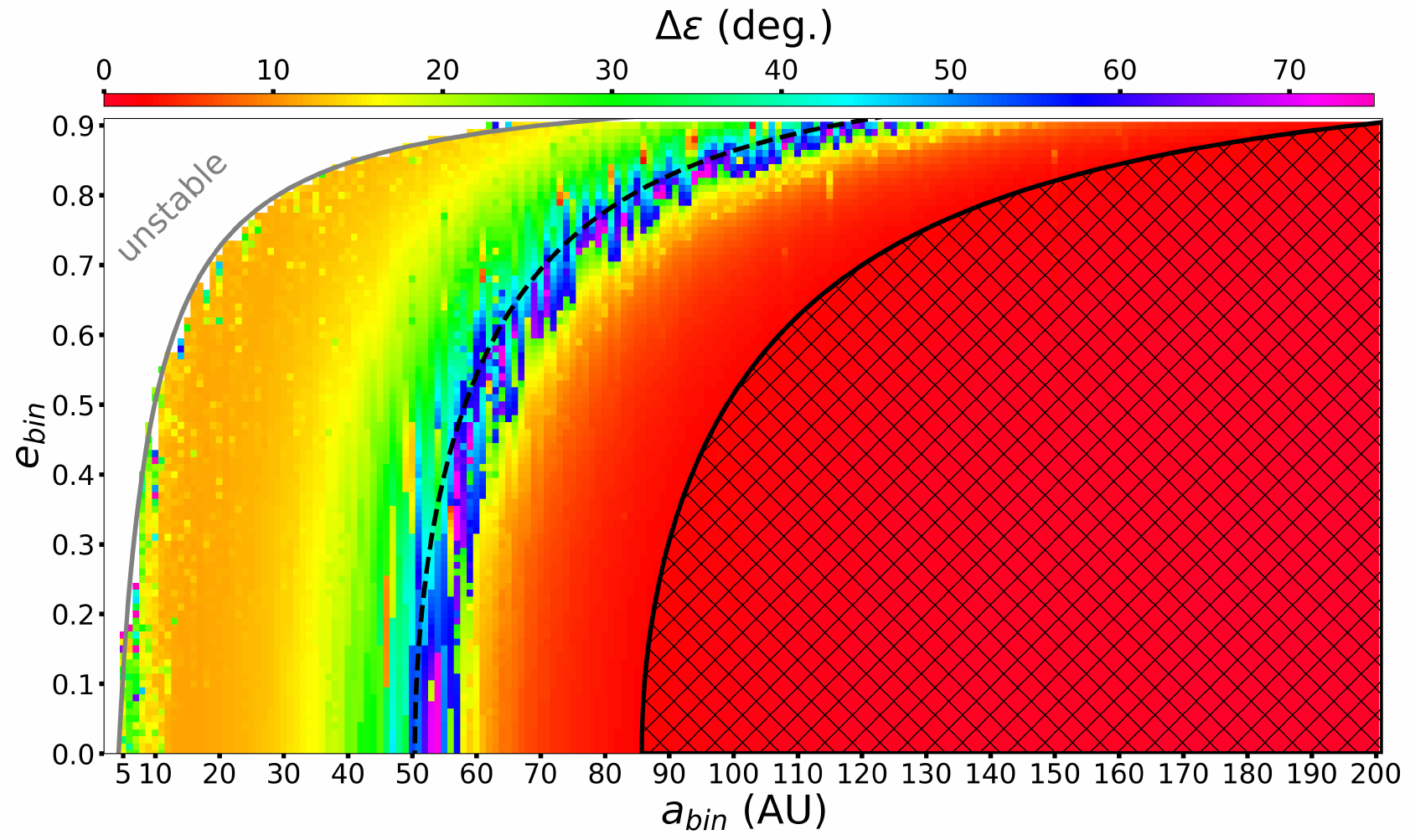}
    \caption{Obliquity variation $\Delta \epsilon$ (color-coded) for a \emph{single} Earth-like rotator ($P_{rot}=23.934$ hr \& $\epsilon_o = 23.4^\circ$) orbiting the primary star at the inner edge of the host star's habitable zone, while the semimajor axis $a_{bin}$ and eccentricity $e_{bin}$ of the binary orbit is varied.  The solid gray curve denotes the boundary between stable (colored) and unstable (white) initial conditions.  The dashed black curve marks the location of a spin-orbit resonance and the hatched black region signifies the regime where the secular orbital precession  frequency is $<20\%$ of the initial spin precession frequency so that $\Delta \epsilon \leq 2.4^\circ$.}
    \label{fig:single_prim_orb}
\end{figure}

Figure \ref{fig:single_prim_orb} demonstrates the obliquity variations (color-coded) of an Earth-like rotator ($P_{rot}=23.934$ hr, $i_p = 5^\circ$, \& $\epsilon_o = 23.4^\circ$) from changing the shape of the binary orbit using initial conditions from \emph{Set 2} and using $\alpha$ Cen-like stellar masses (Section \ref{sec:bin_set2}).  The Earth-like planet begins in an inclined, near circular orbit at the inner edge of the conservative habitable zone of the primary star (M$_A = 1.133$ M$_\odot$ \& L$_A = 1.519$ L$_\odot$).  The irregularities in Fig. \ref{fig:single_prim_orb} are due to the random assignment of the initial spin longitude $\psi_o$. The white area above the gray curve marks which binary parameters destabilize the orbit of a planet.  The black hatched region denotes when {low} obliquity variations occur ($\Delta \epsilon \leq 2.4^\circ$) and the obliquity variations for $a_{bin} > 250$ AU are largely negligible.  In this case, the initial spin precession frequency of the planet is $\sim 9-10$\arcsec/yr while the orbital precession frequency for an $\alpha$ Cen-like binary is $\sim -146$\arcsec/yr.  Using Equation \ref{eqn:sec_freq}, the binary orbit parameters ($a_{bin}$ \& $e_{bin}$) are located (dashed black curve) that allow for overlap between these two precession frequencies and subsequent increases in the obliquity variation. The gradient in obliquity variation for $a_{bin}<40$ AU in Fig. \ref{fig:single_prim_orb} is from the planet scanning the frequency space until the dominant mode is reached and after surpassing the dominant mode the precession frequencies quickly diverge which allows for strong coupling between the spin and orbit angular momentum vectors (similar to Fig. \ref{fig:FFT_aCenB}). 

\subsubsection{Probabilities of Oblquity Variation Using the Binary Star Population}
\begin{figure}
    \centering
    \includegraphics[width=\linewidth]{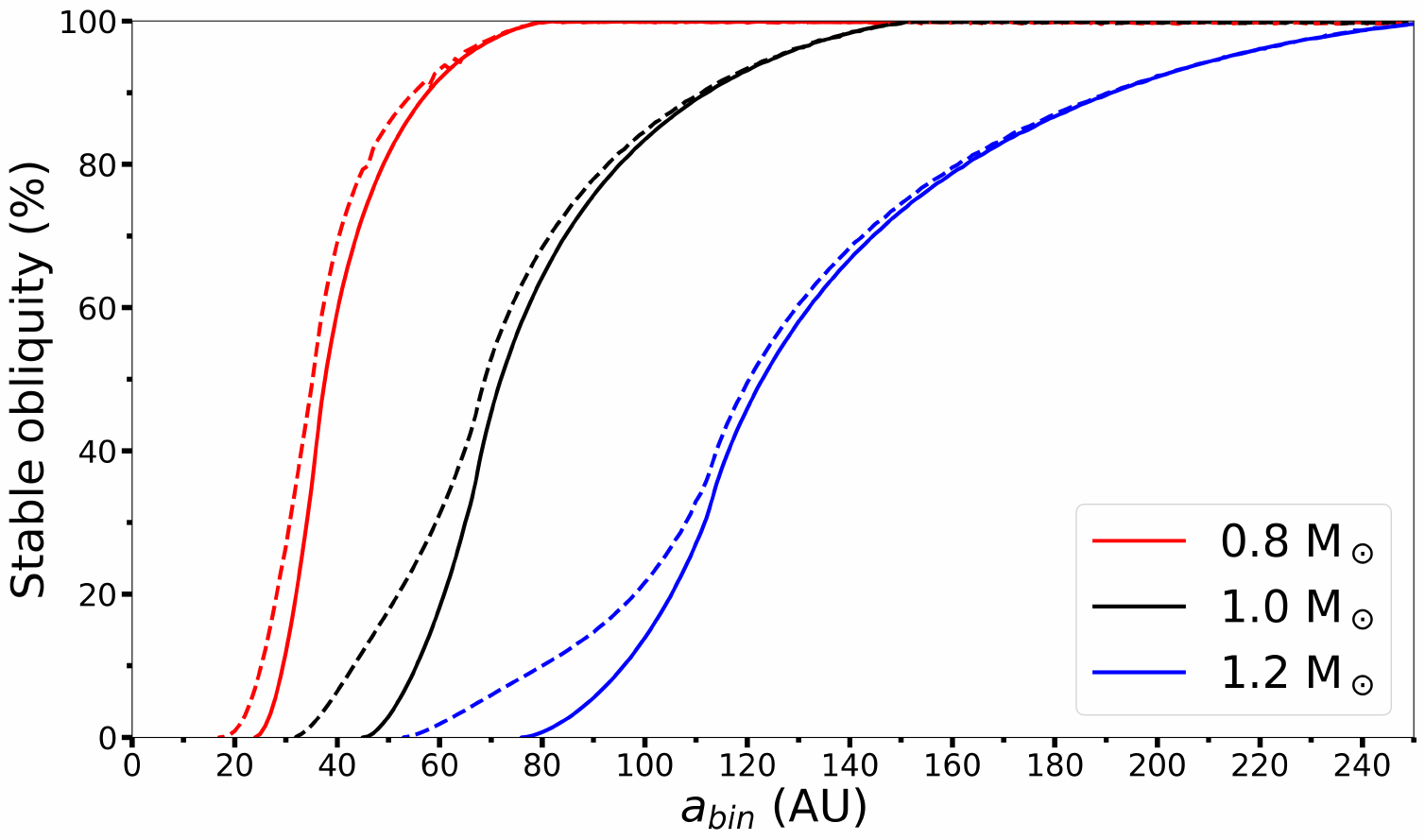}
    \caption{Probability using Monte-Carlo integration for Earth-like stable obliquity ($\Delta \epsilon \leq 2.4^\circ$) considering 0.8 M$_\odot$ (red), 1.0 M$_\odot$ (black), and 1.2 M$_\odot$ (blue) primary stars.  The probability is calculated considering the observed statistics \citep{Moe2017} for binaries with mass ratio quotients between $0.3\leq q \leq 1.0$ (solid) and $0.1\leq q \leq 1.0$ (dashed), where the integration also takes the empirical binary eccentricity distribution into account.  The semimajor axis for the planet is adjusted using the mass-luminosity relation ($L \propto M^4$) so that it begins at the inner edge of the conservative habitable zone. }
    \label{fig:prob_var}
\end{figure}

Figures \ref{fig:single_orb} \& \ref{fig:single_prim_orb} show that median obliquity variation can be estimated for low inclination orbits ($i_p \leq 30^\circ$) for Sun-like stellar binaries and binary separation $a_{bin}$ can be used as a proxy for when Earth-like benign obliquity variations are possible.  We use empirical power laws of binary star populations with Solar-Type primaries (M$_A = 0.8 - 1.2$ M$_\odot$; \cite{Moe2017}) to calculate the probability of stable obliquity ($\Delta \epsilon \leq 2.4^\circ$) for an Earth-like planet orbiting at the inner edge of the host star's conservative habitable zone (see Section \ref{sec:meth_binpop}).  Figure \ref{fig:prob_var} shows the results (color-coded by the primary star mass) of our Monte Carlo calculations, where two ranges for the mass quotient $q$ (=M$_B$/M$_A$) are used (dashed \& solid curves).  The estimates for the dashed curves use a broader range in $q$ that include smaller secondary companions and the power law slope ($0.3\pm0.4$) is not well constrained.  The solid curves provide estimates with a power law slope with better constraints ($-0.5\pm0.3$) and thus are likely to be more reliable.  The general trend in Fig. \ref{fig:prob_var} is that lower mass primary stars will have even lower mass secondary stars and thus, the chance for stable Earth-like obliquity variations is $>50\%$ for $a_{bin} \gtrsim 40$ AU.  For Sun-like (1.0 M$_\odot$) and more massive (1.2 M$_\odot$) primaries, the binary semimajor axis grows to larger separations for a $>50\%$ chance ($a_{bin} \gtrsim 70$ AU and $a_{bin} \gtrsim 120$ AU, respectively).  Given the power law distribution for binary periods, we calculate the percentage of systems where an Earth-like rotator would experience {small ($<2.4^\circ$)} obliquity variations is 87\%, 74\%, or 54\%, depending on the mass of the primary (0.8, 1.0, or 1.2 M$_\odot$ Solar-Type stars, respectively).

\section{Summary \& Conclusions} \label{sec:conclusions}
The obliquity evolution of a single Earth-like planet orbiting within the habitable zone of a Sun-like stellar binary (e.g., $\alpha$ Cen AB) strongly depends on the secular orbital precession induced by the binary companion.  The frequency of the orbital precession $g_s$ can overlap with the spin precession frequency $\alpha\:\cos(\epsilon)$ of the planet, which leads to a larger obliquity variation $\Delta \epsilon$ that is either larger or proportional to twice the planetary mutual inclination with the binary orbit ($\Delta \epsilon \approx 2i_p$).  The obliquity variation can be minimized when the planetary spin longitude aligns with longitude of ascending node ($\psi \approx \Omega_p$).  The minimized obliquity variation is due to librations about a fixed point in the phase space of canonical variables and accounts for $\sim 8\%$ of the possible trajectories.  Observations using polarimetry \citep{Monin2006} suggest that disks are typically misaligned ($\sim$10\degree) with the binary plane and we may expect that the typical obliquity variation ($\sim$20\degree) to be higher than what the Earth experiences.

There are other factors that can modulate the obliquity variations, such as a larger planetary semimajor axis, a different binary orbit, nearby terrestrial planets, or a moon.  The luminosities of $\alpha$ Cen A \& B are quite different (1.519 $L_\odot$ and 0.5 $L_\odot$, respectively), which influences the planetary semimajor axis that receives an Earth-like flux and our simulations of terrestrial planet analogs (including solar system-like mutual inclinations) in both systems show that the obliquity variations are dramatically different depending on the host star.  For systems orbiting $\alpha$ Cen A, the obliquity variation is typically $\sim$1\degree{} due to strong coupling between the planets that keeps the Earth-like world in a near planar orbit.  Moreover, the orbital precession period is much longer than the spin precession period, which allows for strong spin-orbit coupling.  If the binary orbit of $\alpha$ Cen was $\sim$2.5 times larger (i.e., $50-60$ AU), then the orbital precession period would be comparable to the spin precession period and large obliquity variations could ensue.  We combine our results with population studies of binary stars \citep{Raghavan2010,Moe2017} and find the chance that an Earth-like rotator orbiting the primary star would experience {small ($<2.4^\circ$)} obliquity variations is 87\%, 74\%, or 54\%, depending on the mass of the primary (0.8, 1.0, or 1.2 M$_\odot$ Solar-Type stars, respectively).

For systems orbiting $\alpha$ Cen B, the conservative habitable zone is closer to the host star, which allows for a spin-orbit resonance (i.e., $g_s = \alpha\;\cos[\epsilon_o]$) to substantially increase the obliquity variation ($\Delta \epsilon \approx 40^\circ-65^\circ$) for high values of the spin precession constant ($\alpha \gtrsim 60$\arcsec/yr).  An Earth-like spin precession constant ($\sim$46\arcsec/yr) is below this regime, but a moon similar to our own (i.e., Luna-like) would increase the spin precession so that large obliquity variations are common.  This is in contrast to our own Earth-based expectations, where our moon does the opposite \citep{Laskar1993b}.  The amount of spin precession from a moon depends on the moon's mass and semimajor axis, where a Pluto-mass moon at a Luna-like semimajor axis $a_{luna}$ would have a negligible effect.  A smaller semimajor axis (0.2 $a_{luna}$) would allow a Pluto-mass moon to increase the spin precession and allow for larger obliquity variations.  The degree to which a moon can increase the overall spin precession depends on many factors, where they are neither needed nor necessarily even desirable to obtain relatively low obliquity variations.

The obliquity variations in each of these scenarios for $\alpha$ Cen B are important because a planet's obliquity relates to the presence of seasonal cycles and precession determines more broadly how the seasons change over time to influence climatic trends.  Studies of Mars \citep{Head2003,Head2005,Mansfield2018,Kite2019} suggest a combination of changes in atmospheric pressure along with large obliquity variations are responsible for the dramatic difference from Earth-like conditions.  \cite{Forgan2016} showed using a latitudinal energy balance model that Milankovitch cycles would occur on a shorter timescale considering a single Earth-like planet on a near coplanar (0.5\degree) orbit around $\alpha$ Cen B.  However, these variations were mainly due to secular forcing of eccentricity \cite{Andrade2016,Quarles2018a}.  We examine the flux variations (as a proxy for Milankovitch cycles) of an Earth-like planet orbiting $\alpha$ Cen B under a few scenarios: (1) alone with either $i_p = 10^\circ$ or $i_p = 2^\circ$, (2) nearly coplanar with terrestrial planet neighbors, and (3) similar to two with additional precession from a Luna-like moon.  In the first scenario the frequency of the obliquity oscillations match those calculated by \cite{Forgan2016}, but there is a much larger obliquity variation that produces dramatic epochs where the poles typically receive very little flux over an orbit.  In scenario two, we model a system of 4 terrestrial planet analogs from the solar system with near planar mutual inclinations and find the obliquity variation to produce somewhat Earth-like flux variations.  In scenario three we add a Luna-like moon to scenario two and find that the subsequent flux variations to be more similar to Mars, where the polar caps may dominate potential climate trends due to extended periods of extremely low orbit-averaged flux ($F_{avg}<100$ W/m$^2$ for $\sim$1/3 of a secular cycle).  

A planet has not yet been observationally confirmed orbiting either $\alpha$ Cen A or B, although there are many efforts to do so through new techniques with direct imaging \citep{Bendek2015,Belikov2017a,Belikov2017b,Sirbu2017,Sirbu2018}.  Detecting a planet orbiting the closest Sun-like binary would provide some of the necessary constraints to warrant more detailed climate calculations using software like \texttt{VPlanet} \citep{Barnes2019} or coupling our results to 3D global circulation models with a number of assumptions.  Observations with Kepler and K2 have uncovered a few small exoplanets orbiting one star of a stellar binary \citep{Campante2015,Dupuy2016,Mills2017,Ciardi2018,Feinstein2019}, however those binaries are not Sun-like.  The Transiting Exoplanet Survey Satellite (TESS; \cite{Ricker2016}) is expected to observe a multitude of binary stars \citep{Sullivan2015}.  Observations with TESS along with high precision measurements from Gaia \citep{Gaia2018} could allow a more robust statistical approach to studying the obliquity of planets within Sun-like stellar binaries by uncovering deeper insights into planetary occurrence rates.

\section*{Acknowledgments}
The authors would like to thank Steven Kreyche, Jason Barnes, {Konstantin Batygin, and Stephen Kane} for stimulating conversations and detailed comments that improved the clarity of the manuscript.  B.Q. and J.J.L. acknowledge support from the NASA Exobiology Program, grant \#NNX14AK31G.  This research was supported in part through research cyberinfrastructure resources and services provided by the Partnership for an Advanced Computing Environment (PACE) at the Georgia Institute of Technology.

\newpage
\bibliography{bibliography}
\bibliographystyle{aasjournal}

\end{document}